\documentclass[twocolumn]{aastex631}

\usepackage{multirow}

\usepackage{amsmath}

\defcitealias{denBrok2022}{dB22}
\begin{document}

\title{CO isotopologue-derived molecular gas conditions and CO-to-H$_2$ conversion factors in M51}

\author[0000-0002-8760-6157]{Jakob den Brok}
\affiliation{Center for Astrophysics $\mid$ Harvard \& Smithsonian, 60 Garden St., 02138 Cambridge, MA, USA}
\author[0000-0002-9165-8080]{María~J.~Jiménez-Donaire}
\affiliation{Observatorio Astron\'omico Nacional (IGN), C/ Alfonso XII, 3, E-28014 Madrid, Spain}
\affiliation{Centro de Desarrollos Tecnológicos, Observatorio de Yebes (IGN), 19141 Yebes, Guadalajara, Spain}
\author[0000-0002-2545-1700]{Adam Leroy}
\affiliation{Department of Astronomy, The Ohio State University, 140 West 18th Ave, Columbus, OH 43210, USA}
\author[0000-0002-3933-7677]{Eva Schinnerer}
\affiliation{Max Planck Institute for Astronomy, Königstuhl 17, 69117 Heidelberg, Germany}
\author[0000-0003-0166-9745]{Frank Bigiel}
\affiliation{Argelander-Institut für Astronomie, Universität Bonn, Auf dem Hügel 71, 53121 Bonn, Germany}
\author[0000-0003-3061-6546]{Jérôme Pety}
\affiliation{IRAM, 300 rue de la Piscine, F-38406 Saint Martin d’H\`eres, France}
\affiliation{Sorbonne Université, Observatoire de Paris, Université PSL, École normale supérieure, CNRS, LERMA, F-75005, Paris, France}
\author[0009-0005-2122-0680]{Glen Petitpas}
\affiliation{Massachusetts Institute of Technology, 77 Massachusetts Ave., Cambridge, MA 02139, USA}
\author[0000-0003-1242-505X]{Antonio Usero}
\affiliation{Observatorio Astron\'omico Nacional (IGN), C/ Alfonso XII, 3, E-28014 Madrid, Spain}
\author[0000-0003-4209-1599]{Yu-Hsuan Teng}
\affiliation{Center for Astrophysics and Space Sciences, Department of Physics, University of California San Diego,\\ 9500 Gilman Drive, La Jolla, CA 92093, USA}
\affiliation{Department of Astronomy, University of Maryland, 4296 Stadium Drive, College Park, MD 20742, USA}
\author[0000-0003-3537-4849]{Pedro Humire}
\affiliation{Departamento de Astronomia, Instituto de Astronomia, Geofísica
e Ciências Atmosféricas da USP, Cidade Universitária, 05508-900
São Paulo, SP, Brazil}
\author[0000-0001-9605-780X]{Eric W. Koch}
\affiliation{Center for Astrophysics $\mid$ Harvard \& Smithsonian, 60 Garden St., 02138 Cambridge, MA, USA}
\author[0000-0002-5204-2259]{Erik Rosolowsky}
\affiliation{4-183 CCIS, University of Alberta, Edmonton, Alberta, T6G 2E1, Canada}
\author[0000-0002-4378-8534]{Karin Sandstrom}
\affiliation{Department of Astronomy \& Astrophysics, University of California San Diego, 9500 Gilman Drive, La Jolla, CA 92093, USA}
\author[0000-0001-9773-7479]{Daizhong Liu}
\affiliation{Purple Mountain Observatory, Chinese Academy of Sciences, 10 Yuanhua Road, Nanjing 210023, China}
\author[0000-0003-2384-6589]{Qizhou Zhang}
\affiliation{Center for Astrophysics $\mid$ Harvard \& Smithsonian, 60 Garden St., 02138 Cambridge, MA, USA}
\author[0000-0002-9333-387X]{Sophia Stuber}
\affiliation{Max Planck Institute for Astronomy, Königstuhl 17, 69117 Heidelberg, Germany}
\author[0000-0002-5635-5180]{Mélanie Chevance}
\affiliation{Instit\"ut  f\"{u}r Theoretische Astrophysik, Zentrum f\"{u}r Astronomie der Universit\"{a}t Heidelberg, Albert-Ueberle-Strasse 2, 69120 Heidelberg, Germany}
\affiliation{Cosmic Origins Of Life (COOL) Research DAO, \url{coolresearch.io}}
\author[0000-0002-5782-9093]{Daniel A. Dale}
\affiliation{Department of Physics and Astronomy, University of Wyoming, Laramie, WY 82071, USA}
\author[0000-0002-1185-2810]{Cosima Eibensteiner}
\altaffiliation{Jansky Fellow of the National Radio Astronomy Observatory}
\affiliation{NRAO, 520 Edgemont Road, Charlottesville, VA 22903}
\author[0009-0002-9819-1468]{Ina Galić}
\affiliation{Argelander-Institut für Astronomie, Universität Bonn, Auf dem Hügel 71, 53121 Bonn, Germany}
\affiliation{Max Planck Institute for Radio Astronomy, Auf dem Hügel 69, 53121 Bonn, Germany}
\author[0000-0001-6708-1317]{Simon C. O. Glover}
\affiliation{Instit\"ut  f\"{u}r Theoretische Astrophysik, Zentrum f\"{u}r Astronomie der Universit\"{a}t Heidelberg, Albert-Ueberle-Strasse 2, 69120 Heidelberg, Germany}
\author[0000-0002-1370-6964]{Hsi-An Pan}
\affiliation{Department of Physics, Tamkang University, No.151, Yingzhuan Road, Tamsui District, New Taipei City 251301, Taiwan} 
\author[0000-0002-0472-1011]{Miguel Querejeta}
\affiliation{Observatorio Astron\'omico Nacional (IGN), C/ Alfonso XII, 3, E-28014 Madrid, Spain}
\author[0000-0002-0820-1814]{Rowan J. Smith}
\affiliation{SUPA, School of Physics and Astronomy, University of St Andrews, North Haugh, St Andrews, KY16 9SS}
\author[0000-0002-0012-2142]{Thomas G. Williams}
\affiliation{Sub-department of Astrophysics, Department of Physics, University of Oxford, Keble Road, Oxford OX1 3RH, UK}
\author[0000-0003-1526-7587]{David J. Wilner}
\affiliation{Center for Astrophysics $\mid$ Harvard \& Smithsonian, 60 Garden St., 02138 Cambridge, MA, USA}
\author[0009-0007-2660-7635
]{Valencia Zhang}
\affiliation{Center for Astrophysics $\mid$ Harvard \& Smithsonian, 60 Garden St., 02138 Cambridge, MA, USA}
\affiliation{Phillips Academy, Andover, MA 01810, USA}


\begin{abstract}

    Over the past decade, several millimeter interferometer programs have mapped the nearby star-forming galaxy M51 at a spatial resolution of ${\le}170$\,pc. This study combines observations from three major programs: the \emph{PdBI Arcsecond Whirlpool Survey} (PAWS), the SMA M51 large program (SMA-PAWS), and the \emph{Surveying the Whirlpool at Arcseconds with NOEMA} (SWAN). The dataset includes the (1-0) and (2-1) rotational transitions of $^{12}$CO, $^{13}$CO, and C$^{18}$O isotopologues.   {The observations cover the $r{<}\rm 3\,kpc$ region including center and part of the disk, thereby ensuring strong detections of the weaker $^{13}$CO and C$^{18}$O lines.} All observations are convolved in this analysis to an angular resolution of 4$\arcsec$, corresponding to a physical scale of ${\sim}$170 pc. We investigate empirical line ratio relations and quantitatively evaluate molecular gas conditions such as temperature, density, and the CO-to-H$_2$ conversion factor ($\alpha_{\rm CO}$). We employ two approaches to study the molecular gas conditions: (i) assuming local thermal equilibrium (LTE) to analytically determine the CO column density and $\alpha_{\rm CO}$, and (ii) using non-LTE modeling with \texttt{RADEX} to fit physical conditions to observed CO isotopologue intensities. We find that the $\alpha_{\rm CO}$ values  {in the center and along the inner spiral arm} are $\sim$0.5 dex (LTE) and ${\sim}$0.1 dex (non-LTE) below the Milky Way inner disk value. The average non-LTE $\alpha_{\rm CO}$ is $2.4{\pm}0.5$ M$_\odot$ pc$^{-2}$ (K km s$^{-1}$)$^{-1}$. While both methods show dispersion due to underlying assumptions, the scatter is larger for LTE-derived values. This study underscores the necessity for robust CO line modeling to accurately constrain the molecular ISM's physical and chemical conditions in nearby galaxies.

\end{abstract}


%
%
\section{Introduction} \label{sec:intro}

Rotational transitions of CO isotopologues remain one of the most accessible ways to trace the distribution, dynamics, and conditions of the bulk molecular gas in extragalactic studies of the nearby Universe. The molecular gas forms the reservoir for star formation, which occurs primarily within cold, dense Giant Molecular Clouds (GMCs) that are distributed throughout the galaxy, forming part of the interstellar medium (ISM). The low-$J$ transitional lines of $^{12}$CO, the most abundant isotopologue, can be used to trace the general distribution and kinematics of galaxies, but they remain optically thick over most parts of the galaxy. In contrast, due to significantly lower abundances, other CO isotopologues remain optically thin, and their emission is generally an order of magnitude, or more, fainter than bright $^{12}$CO line. After $^{12}$CO, the next most abundant isotopologues are $^{13}$CO and C$^{18}$O \citep{Wilson1994,Henkel2014}. As a result, extragalactic studies have focused on low-$J$ emission of these two CO isotopologue species, which make it possible to trace the entire column of CO-emitting molecular gas \citep[e.g.,][]{Paglione2001,Pan2015,JDonaire2017,Brown2019, Israel2020}. By assuming optically thin emission for $^{13}$CO and C$^{18}$O and comparing their intensities with the optically thick $^{12}$CO lines, we can obtain constraints on the optical depth of the $^{13}$CO or C$^{18}$O lines and relate them to variations in the molecular gas column densities \citep[e.g.,][]{Young1982,Pineda2008, Cormier2018}. Moreover, using multiple rotational-$J$ emission lines for several CO isotopologue lines, it is possible to use radiative transfer models to infer the molecular gas conditions, such as temperature, density, and opacity \citep[e.g.,][]{Goldsmith2008,Israel2020, Teng2022, Liu2023,Teng2023}. Consequently, CO isotopologue line ratios have become a strong diagnostic tool for evaluating the chemical (e.g., abundances) and physical (e.g., temperature, density) properties of molecular gas. A challenge, however, remains that without a sufficient number of rotational-$J$ line observations, studies of line ratios are affected by degeneracies as differences in relative abundance, beam filling factors, opacities, and changes in temperature and density will all affect the measured ratios. 

Initially, our knowledge on the variation of the $^{13}$CO and C$^{18}$O line emission -- and subsequently their abundance ratio -- across the ISM was limited to the Milky Way \citep[e.g.][]{Langer1990, Wilson1994}. In extragalactic studies, at first the focus was on bright, actively star-forming systems where higher column densities and warmer gas made it possible to detect faint CO isotopologues \citep[e.g.][]{Meier2004,Costagliola2011, Aladro2013}. Such systems include (ultra)luminous infrared galaxies ((U)LIRGs), starburst galaxies, or galaxy centers in general. More recently, studies on kpc-scales have provided us with insights into the variation of the various CO isotopologue ratios across normal (i.e., main-sequence) star-forming galaxies and their relation to global galactic properties \citep[e.g.][]{Davis2014, JDonaire2017,denBrok2022,Cao2023}.  These studies report, for instance, an increasing $^{13}$CO/C$^{18}$O intensity ratio with increasing galactocentric radius and a decrease with increasing star formation rate surface density \citep{JDonaire2017}, or increasing $^{12}$CO/$^{13}$CO intensity ratios with the star formation rate surface density \citep{Davis2014, Cao2023}. From these studies alone, it remains unclear to what degree these observed ratio variations are the result of changes in the opacity due to different densities of molecular gas or changes in the relative abundances due to physical or chemical processes that affect the amount of the CO isotopologue species. Studies that focus on numerous rotational-$J$ transitions to break the aforementioned degeneracy to some degree are still limited mainly to the bright centers of nearby star-forming galaxies \citep[e.g.][]{Israel2020, Teng2022, Teng2023}.

\begin{figure}
    \centering
    \includegraphics[width=\columnwidth]{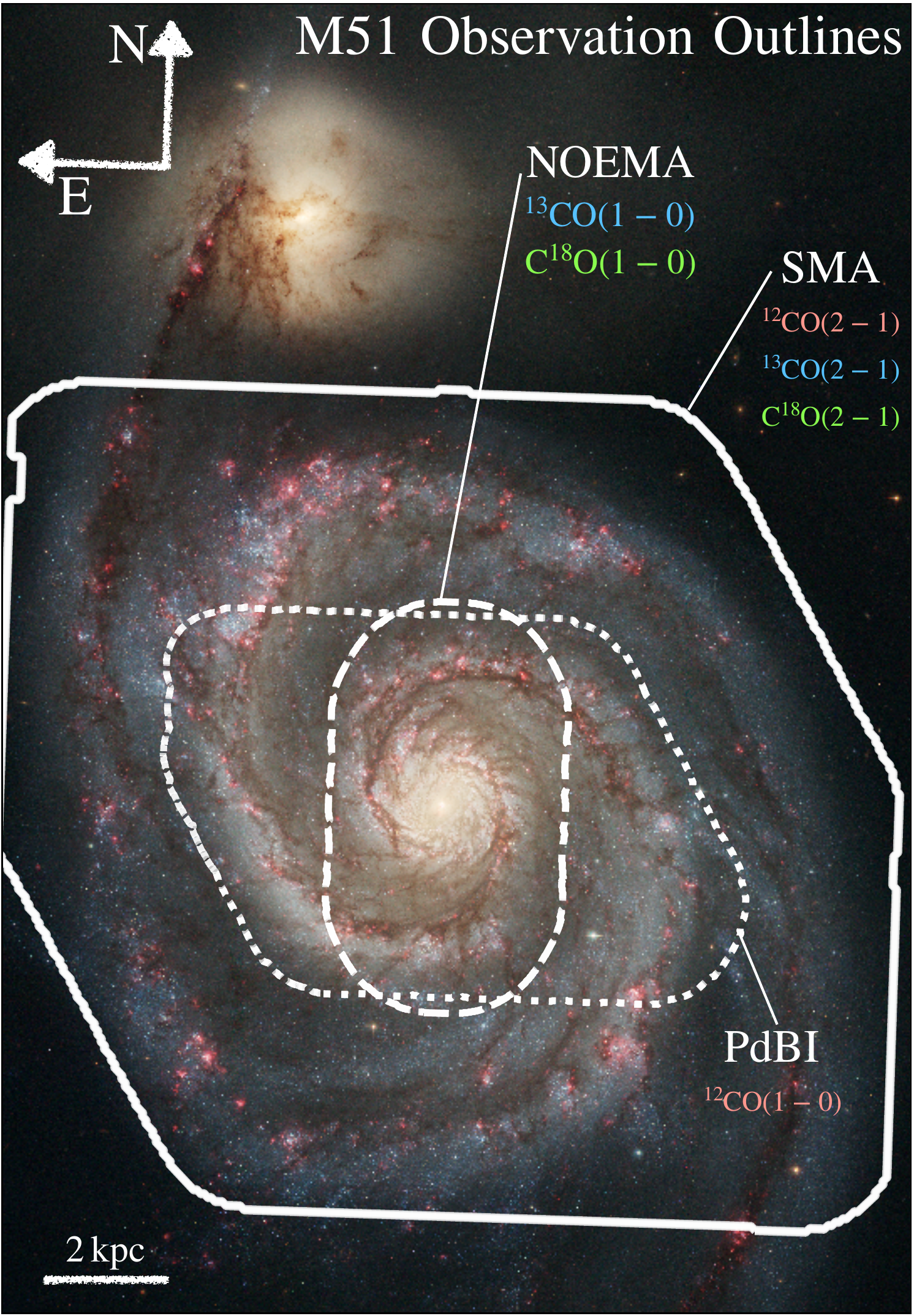}
    \caption{{\bf Spatial Coverage of Observations} In this study, we rely on Plateau de Bure Interferometer (PdBI), Northern Extended Millimeter Array (NOEMA), and Submillimeter Array (SMA) observations, each of which covered some extent of the disk of M51. The boxes indicate which lines were observed by each respective program. Background HST image credit: NASA, ESA, S.~Beckwith (STScI), and The Hubble Heritage Team (STScI/AURA).}
    \label{fig:outline}
\end{figure}

Given a set of rotational-$J$ CO (isotopologue) lines, a common approach in the literature is to assume local thermodynamic equilibrium (LTE; e.g., \citealt{Wilson2009}), which makes it possible to solve the equations of radiative transfer analytically and infer the molecular gas conditions. This technique is generally (computationally) easy and fast to implement and only requires the detection of two emission lines (optically thick $^{12}$CO and optically thin $^{13}$CO) to constrain the excitation temperature, the opacity, and the CO column density. Therefore, this is often the preferred approach for extragalactic studies of nearby galaxies \citep[e.g.,][]{Rigopoulou1996,Dahmen1998,Paglione2001,Braine2010,Davis2014,Cormier2018,MorokumaMatsui2020, Cao2023}, where multi-CO line surveys are challenging to achieve due to the required sensitivity of the observations. However, the underlying assumption that the level populations are following thermal equilibrium is often not accurate in the molecular interstellar medium \citep[e.g.,][]{Shirley2015}. Therefore, non-LTE  {software tools} have been developed, which  {can} solve the radiative transfer equations numerically and do not require the level populations to be in thermal equilibrium.  {Such} an implementation that models line intensities for a specified set of physical conditions is \texttt{RADEX} \citep{vanTak2007}. To obtain robust constraints, multiple CO lines have to be observed to limit the number of free parameters. Therefore, non-LTE approaches are not only computationally, but also observationally more expensive. Consequently, extragalactic studies employing the non-LTE approach are generally limited to focusing on bright galaxy centers \citep[e.g.,][]{Israel2020, Teng2022, Teng2023} or brighter starburst or merging galaxies \citep[e.g., ][]{Salak2014,He2024}. With our dataset of M51 at hand, which contains several CO isotopologue lines, we can compare the use of an LTE and non-LTE approach simultaneously to contrast the two methods and constrain key parameters, such as CO column density and the CO-to-H$_2$ conversion factor in the context of regular galaxy conditions beyond the center or starburst regime.
{So far, these LTE and non-LTE multi-line modeling efforts for extragalactic observations often rely on simplistic one- or two-zone models to describe the molecular gas distribution due to limitations imposed by the number of detected lines and the angular resolution.  A one zone description is, however, generally only a poor representation of the actual underlying (sub-beam) gas distribution. For instance, turbulent theories, supported by Milky Way observations that resolve individual clouds \citep[e.g.,][]{Goldsmith2008, Kainulainen2009, Tafalla2023}, predict a range of densities that follow a lognormal distribution within each cloud \citep[][]{Padoan2002}. \citet{Leroy2017dens} have investigated how the sub-beam density distribution affects the beam-averaged emissivity in the context of extragalactic observations. Their findings suggest that the beam-averaged emissivity of dense gas tracers (e.g., HCN, HCO$^+$) is highly sensitive to variations in sub-beam density distributions, whereas the low-J CO line emission is relatively insensitive to these variations. Therefore, while a single- or two-component CO line modeling approach should be interpreted with care, it can still be viable for capturing the bulk gas properties.
}
This study focuses on the bright face-on nearby grand-design spiral galaxy M51 (NGC 5194). It is at a distance of D ${\approx}$ 8.6 Mpc \citep{McQuinn2016} – so an angular resolution of 4 arcsec translates to a physical resolution of ${\sim}$170 pc. The galaxy interacts tidally with a companion (see \autoref{fig:outline}). The disk is dominated by molecular gas \citep{Schuster2007, Leroy2008}, making it possible to detect also the fainter CO isotopologues across a large part of the star-forming disk of M51. The spiral arms are particularly prominent in observations across the electromagnetic spectrum. Also, the molecular gas is clearly distributed along these spiral arms \citep{Koda2011,Pety2013,Schinnerer2013}. In a previous IRAM 30m large program study \citep[CLAWS;][hereafter \citetalias{denBrok2022}]{denBrok2022}, the CO isotopologue line emission of $^{12}$CO, $^{13}$CO, and C$^{18}$O across the disk at kpc-scales has been studied. Galaxy-wide, the various CO isotopologue line ratios investigated there show trends in agreement with selective nucleosynthesis and changes in optical depth as main drivers. Here, we follow up this study by investigating the same line ratios at higher physical resolution to obtain a more detailed view of the potential drivers of CO isotopologue line ratio variation.

This paper is structured as follows: We describe the various datasets used in Section \ref{sec:obs}. In Section \ref{sec:results}, we present the main results of this study, including the line ratio trends across  {the central $r{<}3\,kpc$ region of} M51 and the variation in molecular gas properties derived from radiative transfer calculations. We discuss the interpretation of the results in Section \ref{sec:disc}. We conclude in Section \ref{sec:conc}.

%
%

\begin{deluxetable*}{ccccccc}

\tablecaption{Summary of the CO isotopologue line observations relevant in this study.}
\label{tab:obs_lines}

\tablehead{\colhead{Line} & \colhead{$\nu_{\rm rest}$} & Native Beam\tablenotemark{a} & \colhead{$\langle$sensitivity$\rangle$\tablenotemark{b}} & \colhead{Completeness\tablenotemark{c}}&\colhead{Telescope/Survey/Reference}  \\ 
\colhead{} & \colhead{[GHz]} & \colhead{[$\arcsec$]} & \colhead{[mK]} & \colhead{[\%]} } 

\startdata 
         C$^{18}$O(1-0)&109.782&$2.5{\times}2.2$&14&32& 
         NOEMA / SWAN\\
         $^{13}$CO(1-0)&110.201&$2.4{\times}2.1$&17 & 63 & Stuber et al. (in prep.); see also \citet{Stuber2023} \\
          \hline
         $^{12}$CO(1-0)&115.271&$1.1{\times}1.1$ &105&100& PdBI / PAWS / \citet{Schinnerer2013}\\
          \hline
         C$^{18}$O(2-1)&219.560&$4.1\times4.1$&55&2&\multirow{3}{*}{SMA / SMA-PAWS / this study; see \autoref{app:dataReduxSMA}}\\
         $^{13}$CO(2-1)&220.399&$4.1\times4.1$&70&14\\
         $^{12}$CO(2-1)&230.538&$3.9\times3.9$&90&73\\ 
\enddata

\tablecomments{(${\rm a}$) The beam size of the original data. For the subsequent analysis, we convolve all cubes to a common beam size of $4''$. ($\rm{b}$) The median rms of the data convolved to $4''$, per 5\,km\,s$^{-1}$ channel width. (c) Percentage of sightlines (i.e. spatial pixels) with respect to $^{12}$CO(1-0) within the NOEMA FoV that have significant detection (per definition, it is 100\% for \mbox{$^{12}$CO(1-0)}).}

\end{deluxetable*}

\section{Data and Observations} \label{sec:obs}
Here, we use observations of the (1-0) and (2-1) rotational transitions of $^{12}$CO, $^{13}$CO, and C$^{18}$O from two different mm interferometers. The outlines in \autoref{fig:outline} illustrate the spatial extent of the different observations. We provide a summary of the CO isotopologue line observations in Table \ref{tab:obs_lines}. We present all spectra in terms of the line-of-sight velocity, $v_{\rm LSR}$, relative to the local standard of rest.

\subsection{Plateau de Bure Interferometer -- \texorpdfstring{$^{12}\rm CO$}{Lg}(1-0)}
The \textit{PdBI Arcsecond Whirlpool Survey} (PAWS) observed the $^{12}$CO(1-0) line emission at $1$\,arcsec ($\approx 40$\,pc) resolution across the center and a large part of the disk of M51 \citep{Schinnerer2013}. The precise field-of-view of the observation is highlighted in \autoref{fig:outline}. The observations of the Plateau de~Bure Interferometer (PdBI) have been short-spacing corrected using IRAM \mbox{30-m} single-dish data. The data reduction is described in \citet{Pety2013}.

\subsection{Northern Extended Array -- \texorpdfstring{$^{13}\rm CO$ and $\rm C^{18}O$}{Lg}(1-0)}

The \textit{Surveying the Whirlpool at Arcseconds with NOEMA} (SWAN) is an IRAM large program (LP003 PIs: E. Schinnerer, F. Bigiel; \citealt{Stuber2023}). Observations were carried out using the Northern Extended Millimetre Array (NOEMA) and IRAM 30m single dish for short-spacing corrections. The observations targeted several molecular lines in the 3 mm line window within the central $5{\times}7$ kpc of M51 (see the outline in \autoref{fig:outline}). The spectral coverage of the program includes the 1-0 transitions of $^{13}$CO and C$^{18}$O. The data have a  {native} angular resolution of  {${\sim}2.5''$ (${\sim}$100\,pc)}.
For details about the data calibration and reduction, we refer the reader to S. Stuber et al. (in prep.).

\subsection{Submillimeter Array -- \texorpdfstring{$^{12}\rm CO$, $^{13}\rm CO$ and $\rm C^{18}O$}{Lg}(2-1)}

As part of a Submillimeter Array (SMA) Large Program (2016B-S035, PI: K. Sliwa), the molecular disk of M51 was observed in three configurations: subcompact (SUB), compact (COM), and extended (EXT). While the SUB configuration consists of 147 pointings, covering the entire molecular disk of M51, the observations in COM consist of 55 pointings covering the same footprint as the PAWS data. The SMA observations include the $^{12}$CO\,(2-1), $^{13}$CO\,(2-1), and C$^{18}$O\,(2-1) lines.  For the subsequent analysis, we combine the observations in COM and SUB configurations, limiting the footprint to the overlapping field of view (which is equivalent to the PAWS field of view).  Furthermore, the SMA observations have been short-spacing corrected using IRAM 30m observations from the CLAWS large program (\citetalias{denBrok2022}).  We document the details of the data reduction steps in Appendix \ref{app:dataReduxSMA}. The SMA data cubes and associated moment maps have been made publicly available \citep{DVN/IE2CC0_2024}\footnote{\url{https://doi.org/10.7910/DVN/IE2CC0}}.


\subsection{Ancillary Datasets}
In this study, we compare the distribution and derived properties of the molecular gas from the CO isotopologues to the distribution of the star formation rate across M51. For this purpose, we need comparable high angular resolution star formation rate (SFR) tracers. Previous studies that investigated the connection between the molecular gas condition and star formation in M51 mainly relied on IR-based SFR tracers \citep[e.g.,][ \citetalias{denBrok2022}]{Leroy2017}. However, these IR-based tracers are limited in angular resolution to ${\ge}10''$. 

In this study, we use 33\,GHz extended Very Large Array (EVLA) observations of M51 (\citealt{Querejeta2019}) 
to trace the SFR. The angular resolution of this data is ${\sim}3''$.  The 33\,GHz continuum, which mainly traces free-free emission, is empirically calibrated to a measure of SFR \citep{Murphy2011}.  We use the SFR surface density map derived by \citet{Querejeta2019}. For extensive details on the particular method to compute the SFR surface densities, we refer to \citet{Querejeta2019}. In short,  {they estimated} the thermal fraction (i.e., fraction from free-free emission) in individual 100\,pc apertures by contrasting the 33\,GHz emission to other radio bands (including the 1.4\,GHz, 4.9\,GHz, and 8.4\,GHz bands). They report typical thermal fractions of 55-75\% in M51\footnote{This thermal fraction is lower compared to other galaxies, where fractions of ${\sim}90\%$ are reported \citep{Linden2020}. The low thermal fractions in M51 is likely related to  the ongoing merger with its companion galaxy M51b and the fact that the AGN likely also contributes diffuse synchrotron emission}. The free-free luminosity, $L_{33\,\rm GHz}^T$, can then be converted to an SFR estimate using the following empirical relation \citep{Murphy2011}:
\begin{equation}
\small
    \left(\frac{\rm SFR}{M_\odot\,\rm yr^{-1}}\right) = 1.5\times10^{-27}\cdot\left(\frac{T_e}{10^4\,\rm K}\right)^{-0.45}\cdot\left(\frac{L^T_{33\,\rm GHz}}{\rm erg\,s^{-1}\,Hz^{-1}}\right),
\end{equation}
where $T_e$ describes the electron temperature (following \citealt{Querejeta2019}, we assume $T_e=6300\,\rm K$ based on previous observations by \citealt{Croxall2015}). For the calculation, we use a thermal fraction of 75\%.

We note that the VLA data has not been short spacing corrected. Therefore, given the uncertainty of the missing flux (see \citet{Querejeta2019} for details), we emphasize that the SFR measurements represent a lower limit.
We stress, however, as the observations were carried out using the very compact configuration of the VLA interferometer (sensitive to spatial scales up to $44\arcsec {\approx} 1.6\,\rm kpc$), flux filtering effects will be limited. This is further supported by \citet{Murphy2018} who find that the missing flux is not significant when comparing VLA and GBT single-dish observations. 

\subsection{Data Homogenization and Processing}
\begin{figure}
    \centering
    \includegraphics[width=\columnwidth]{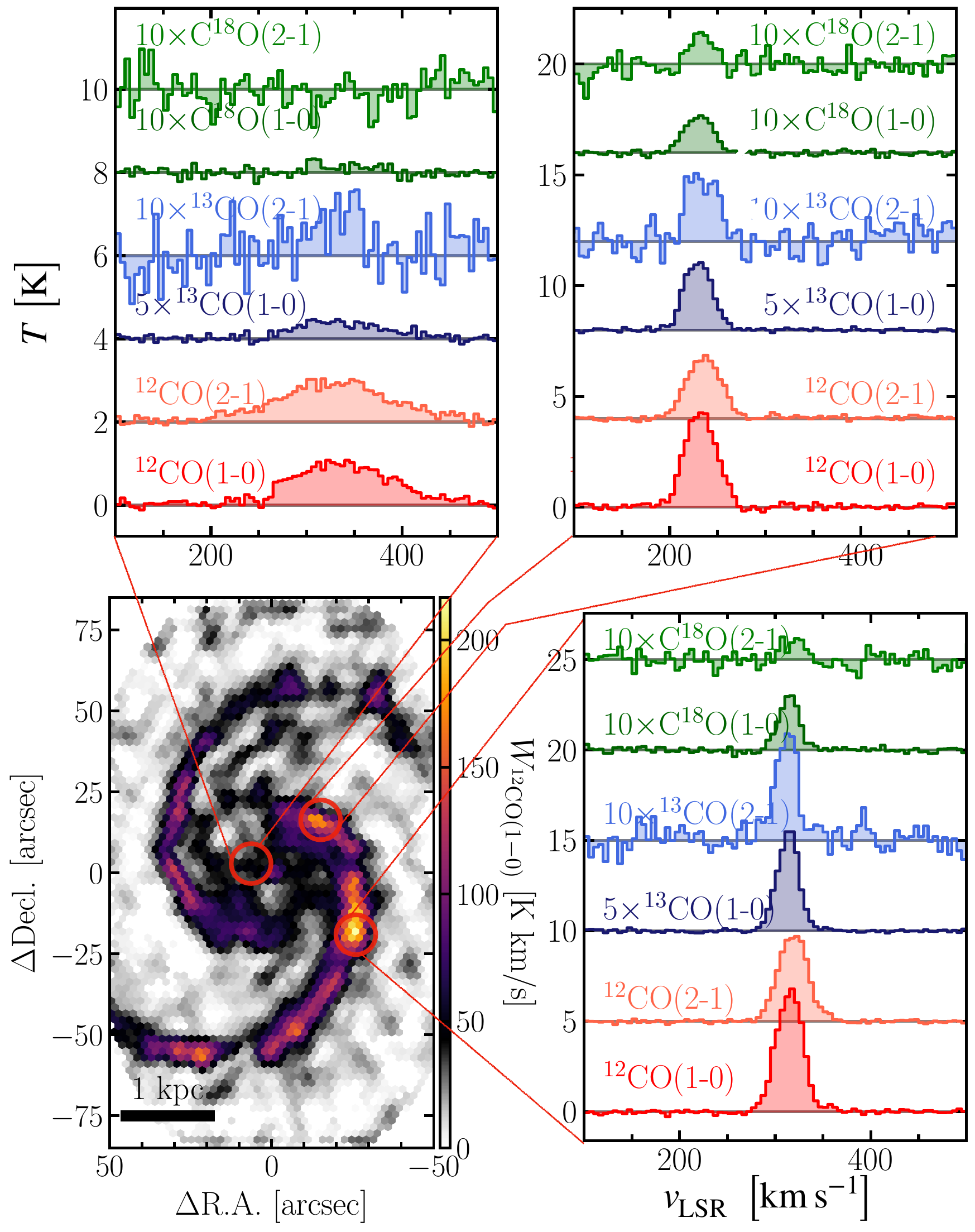}
    \caption{{\bf CO Isotopologue Line Detection} The  {bottom left} panel illustrates the $^{12}$CO(1-0) moment-0 map in units of $\rm K\,km\,s^{-1}$. The top three panels present the line emission spectra for three selected pointings in the galaxy. The spectra are each plotted with an incrementing offset, from the left to the right panel, of $\Delta T_{\rm mb}{=}$5\,K, 2\,K, and 4\,K, respectively. We scale the $^{13}$CO and C$^{18}$O spectra by a factor of 5 or 10 such that the line emission is visible when compared to the much brighter $^{12}$CO line.}
    \label{fig:M51_data}
\end{figure}

For a proper analysis combining all the CO isotopologue line observations from various observatories, we use the \texttt{PyStructure}\footnote{\url{https://github.com/jdenbrok/PyStructure/}} pipeline code  {\citep{PyStructure_v3}}. This tool convolves all cubes to a common beamsize and regrids them to a hexagonal grid with 5\,$\rm km\,s^{-1}$ channel width. We convolve all cubes and maps to $4''$ resolution with a Gaussian kernel using the astropy \texttt{convolve} function. 
We resample each cube and map on a hexagonal grid of points, which are half-beam sized separated (i.e.\ $2''$). The grid spans the entire field-of-view of the $^{12}$CO(1-0) observation. The particular steps are done using the \texttt{PyStructure} pipeline presented in \citetalias{denBrok2022}.

The CO isotopologue data cubes are processed in this framework to derive the moment-0 (``\emph{integrated intensity}"), moment-1 (``\emph{weighted sightline velocity}"), and moment-2 (``\emph{line FWHM}") maps\footnote{To convert the FWHM to a Gaussian line width equivalent, $\sigma$, we need to divide the former by the factor 2.355.}. The moment-0 is computed by integrating over a masked spectral range. The range we integrate over is sightline dependent. For this computation, the same velocity range is used for each line. We use the highest-S/N line,  $^{12}$CO(1-0), to determine the spectral range where we expect emission by  {expanding the} high $\rm S/N{>}4$  {mask into the} low $\rm S/N{>}2$ mask,  {such that only those connected voxels at $\rm S/N{>}2$ remain that are associated with a $\rm S/N{>}4$ peak}.  For the moment-1 and moment-2 calculations, we determine a more strict mask for each sightline and for each CO isotopologue line separately. This mask requires that at least 3 consecutive channels are significantly detected above 4$\sigma$ of the rms. To estimate the error and uncertainty, we compute the standard deviation of the line-free part of the spectrum per sightline for each CO isotopologue. This yields the noise per $5\,\rm km\,s^{-1}$ channel width. We propagate the errors assuming independent channels to compute the noise for the different moment maps.

\autoref{fig:M51_data} illustrates the resulting  data set used in the subsequent analysis. The bottom panel shows the moment-0 map of $^{12}$CO(1-0), convolved to $4''$ and resampled on a hexagonal grid. We depict the spectral lines for three selected points within the central region of M51 where we expect significant emission from all six CO isotopologue lines.

\subsubsection{Spectral line stacking}
To improve the signal-to-noise and retrieve trends with key galactic parameters, we employ a spectral line stacking method, which is described in detail in \citet{Schruba2011,Calduprimo2013,Donaire2019}; and \citet{Neumann2023}. In short, spectral stacking consists of  {first velocity-normalizing the spectra (i.e. shifting the profiles such that any velocity offsets due to rotation and/or non-streaming motions relative to the systemic velocity of the galaxy are removed.), then} binning these spectra by a respective quantity, and finally averaging  {the spectra per bin}. This technique exploits the fact that the noise will not add coherently unlike any faint emission present in the data. For more details and an analysis of the performance of the method for interferometric data, we refer to \citet{Neumann2023}.

We stack the data by galactocentric radius, SFR surface density, and FWHM of the $^{12}$CO(1-0) emission line. For galactocentric radius, the stacking bins range from 0 to 3\,kpc in steps of 300\,pc bin-width. For SFR surface density, we use logarithmic bins. We use 10 bins ranging from $\log\left(\Sigma_{\rm SFR}/\left(M_\odot\,\rm yr^{-1}\,kpc^{-2}\right)\right)$ = $-3$ to $0$. For FWHM, we also use logarithmic bins ranging from 8\,km\,s$^{-1}$ to 70\,km\,s$^{-1}$. 

\subsubsection{CO line ratio computation and notation}\label{ratio_comp}
We compute the integrated line intensity ratios (hereafter just ``line ratio") by dividing the moment-0 maps. In our analysis, we refer to line ratios as significant if both lines are detected with a significance of ${\ge}5\sigma$,  {which translates into a ${\gtrsim}4\sigma$ ratio measurement. We note that the number of significant detections (and hence the spatial distribution) varies for each line ratio depending on the sensitivity and strength of the two lines. We illustrate the spatial extent for the significantly detected sightlines per line ratio in Appendix \ref{app:detection_extension}. The ${\ge}5\sigma$ threshold translates into a molecular gas point-source mass sensitivity\footnote{Assuming a line width of 15\,km\,s$^{-1}$ and a conversion factor of 2\,$M_\odot\,\rm pc^{-2}\,(K\,km\,s^{-1})^{-1}$.} for $^{12}$CO(1-0) of ${\sim}1\times 10^6\,M_\odot$}. In this paper, we refer to the line ratios with the following notation. The superscript indicates which CO isotopologue(s) is(are) used. The subscript indicates information about the rotational-$J$ transition. For instance, $R^{13/12}_{10}$ is the $^{13}\rm CO/^{12}CO$(1-0) integrated intensity ratio and $R_{21}^{12}$ indicates the $^{12}\rm CO(2-1)/(1-0)$ ratio\footnote{The line ratio of two different rotational transitions is abbreviated by the subscript as ``21"$:{\bf 2}\rightarrow1/{\bf 1}\rightarrow0$. This is also evident if the superscript only contains one CO isotopologue.}. In case one of the two lines is not detected (i.e., $\rm S/N{<}5$), we compute a limit. If the line in the numerator is not detected, it will be an upper limit. In case the line in the denominator is not detected, we will compute a  lower $5\,\sigma$ limit. The limit is computed using the measured spectral rms and assuming a line width of 20\,km\,s$^{-1}$.

 {We note that the use of sigma-clipping introduces a bias toward CO-bright regions in M51, such as the central area and the onset of the spiral arms. This is particularly important for ratios involving the fainter $^{13}$CO(2-1) and C$^{18}$O(2-1) lines, where both the reduced line strength and the lower sensitivity of the SMA observations limit the detectable signal. In contrast, when stacking, we incorporate all sightlines, regardless of detection significance, reducing the bias toward CO-bright regions. Consequently, stacking may result in different averages compared to a traditional median, as the inclusion of sightlines with non-detections potentially shifts the overall value. Therefore, we view the stacking results as being more robust (less biased) and hence focus, in particular for the derived line ratio trend analysis, on using stacking results.}
%
%
\begin{figure*}
    \centering
    \includegraphics[width=0.96\textwidth]{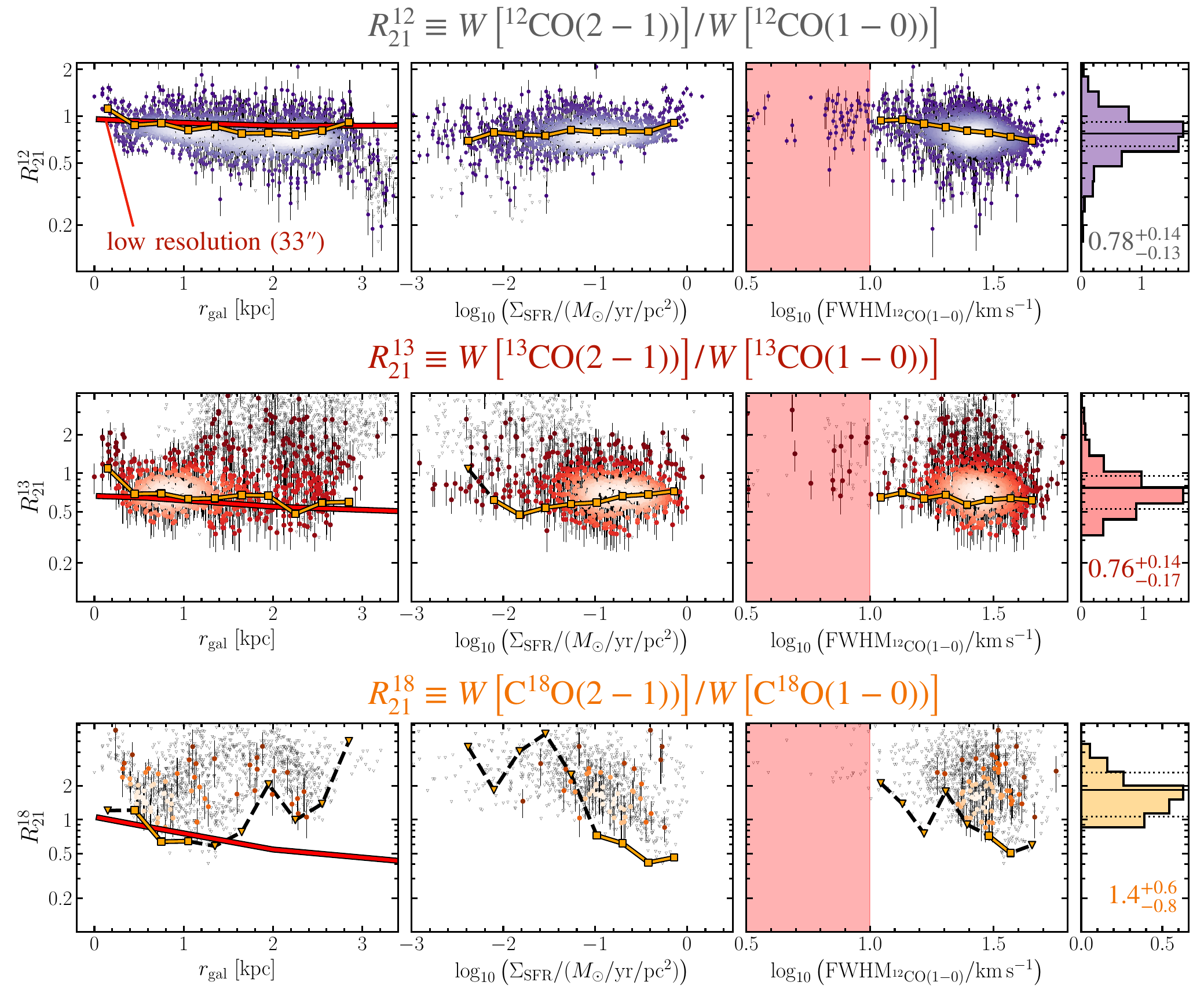}
    \caption{{\bf CO Isotopologue Ratio Trends} These panels show the line ratio as a function of galactocentric radius (left), SFR surface density (middle), and FWHM of $^{12}$CO(1-0) (right) panel. Only lines of sight within the NOEMA field-of-view are considered. We show the individual sightlines where both lines have $\rm S/N{\ge}5$ as colored points. The small black triangles indicate 5$\sigma$ upper limits. In addition, the orange line depicts the stacked line ratio trend. We stack all points for which $^{12}$CO(1-0) is significantly detected. The red line in the left panels is the radial trend obtained at low-angular resolution from \citetalias{denBrok2022}. The red shaded region in the right panels indicates where the FWHM is below two times the channel width of our data (10\,km\,s$^{-1}$). The histograms show the distribution of these significantly detected data points per line ratio. The black line represents the weighted average (weighted by the $^{12}$CO(1-0) intensity) and the dotted lines the weighted 16$^{\rm th}$-to-84$^{\rm th}$ percentile range.}
    \label{fig:rgal_ratio}
\end{figure*}

\begin{figure*}
    \centering
    \includegraphics[width=0.96\textwidth]{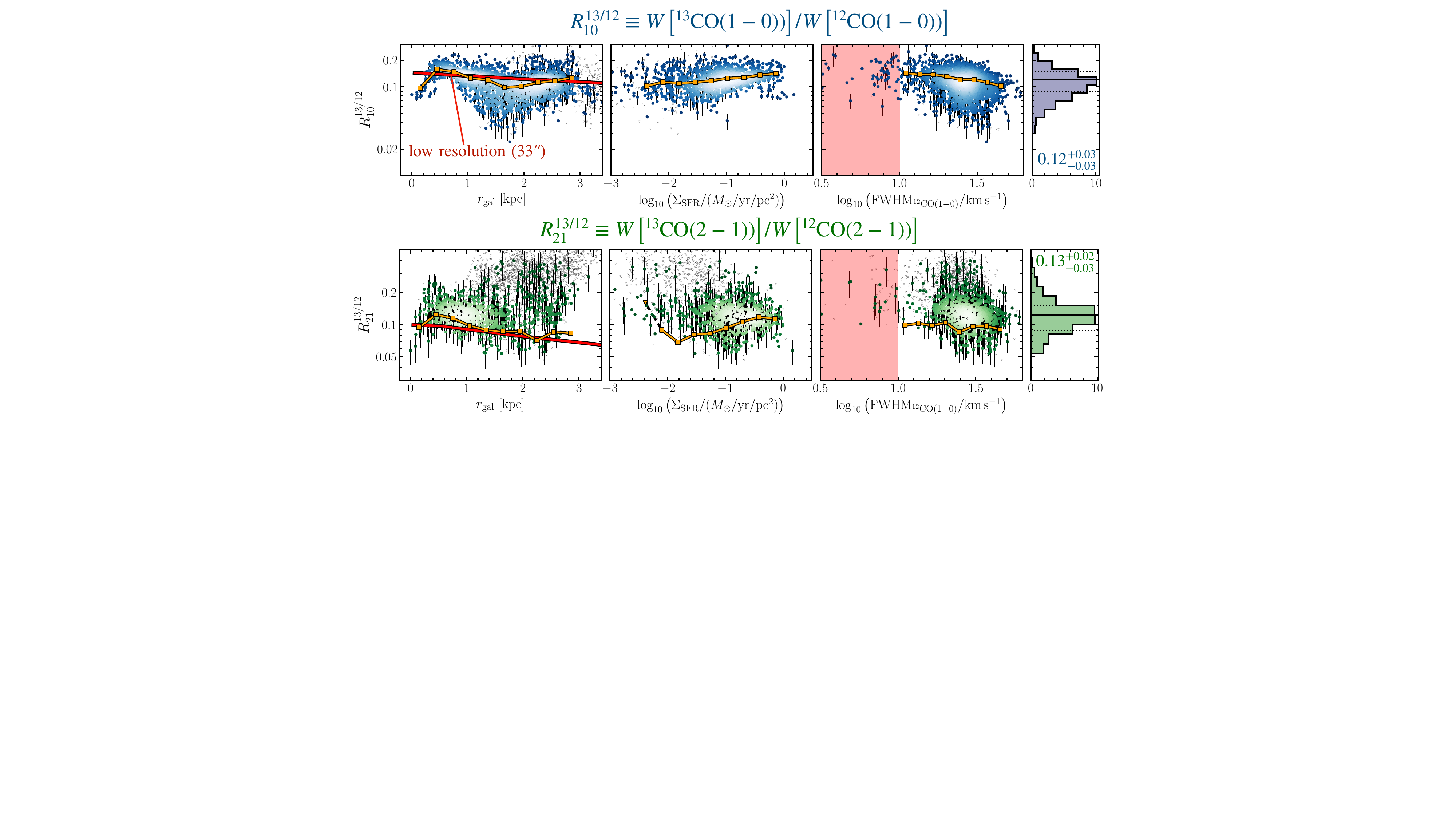}
    \caption{{\bf CO Isotopologue Ratio Trends (continued).} These panels have the same description as in \autoref{fig:rgal_ratio}.}
    \label{fig:rgal_ratio_2}
\end{figure*}

\section{Results} \label{sec:results}
We first focus on the overall CO isotopologue line ratio trends across  {the central $r{<}3\,\rm kpc$ region of} M51. For this purpose, we compute the azimuthally averaged and (galactocentric) radial trends. We also investigate the isotopologue ratio trends with the SFR surface density and the $^{12}$CO(1-0) line FWHM. We then perform an LTE line modeling analysis based on the $^{12}$CO excitation and the $^{12}$CO/$^{13}$CO(1-0) ratios. Finally, we also present results from a non-LTE line modeling approach using \texttt{RADEX}.

\subsection{General CO isotopologue line ratio trends} \label{sec:ratiotrend}
We first investigate the sightlines for which we have significantly detected line ratio values. Earlier studies find such radial trends based on resolved measurements of CO isotopologues and C and O isotopes in the Milky Way \citep[][]{Langer1990,Milam2005,Yan2023} and nearby galaxies \citep[e.g.][]{JDonaire2017, Cormier2018, Martin2019}. We compare these ratios to the SFR surface density, because it relates to and traces to some degree the molecular gas conditions. Particularly, we expect higher SFR surface density to correlate with hotter (heated by young stars) or denser (gas in the process of forming stars) gas \citep{Narayanan2014}. Finally, we also contrast the line ratios to the $^{12}$CO(1-0) line FWHM which we expect to trace besides the larger scale gas motions also, to some limited degree, the turbulence of the gas \citep{Baker1976, Shetty2011}. We also expect that the turbulence of the gas has an impact on the optical depth of the CO emission \citep{Shetty2011}. The trends are shown in \autoref{fig:rgal_ratio} and \autoref{fig:rgal_ratio_2} and relevant correlation coefficients are provided in \autoref{table:results_ratio}.  {If not specifically stated, the trends refer to the stacked line ratio measurements.}

\subsubsection{\texorpdfstring{$R^{12}_{21}$}{Lg} Ratio}
$R^{12}_{21}$ has been thoroughly studied at low (kpc-scale) and high (cloud-scale) angular resolution across and within nearby galaxies \citep[e.g.][]{Koda2012,Brown2021, Yajima2021, Leroy2022, denBrok2023b}. Using our high angular resolution observations (see top panel of \autoref{fig:rgal_ratio}), we report a disk-wide, $^{12}$CO(1-0) intensity weighted line ratio average and 16$^{\rm th}$-to-84$^{\rm th}$ percentile range of $\langle R_{21}^{12}\rangle^{\rm weight.}=0.78^{+0.14}_{-0.13}$. At 33$''$ angular resolution, the line ratio appears flat as a function of galactocentric radius within the central 3\,kpc of the galaxy (see red line in the left panel of \autoref{fig:rgal_ratio}). At $4''$ angular resolution, we observe a flat trend of $R_{21}^{12}$ with an increase towards the central $r{<}500\,$pc radii. The individual sightlines show a sharp decrease from the center to 0.25\,kpc from a value of ${\sim}1.25$ to ${\sim}0.9$, while $R_{21}^{12}$ flattens at $r_{\rm gal}{\ge}0.5\,$kpc (the orange line)\footnote{A note of caution when interpreting the stacked line ratio trends  {(see also explanation in Section \ref{ratio_comp})}: in the various panels, we show the individual sightlines as single data points. For the stacking, however, we also include sightlines where the line emission (apart from $^{12}$CO(1-0)) is insignificant. If, for a given bin, the number of pixels without significant detection dominates, then the resulting line stack will have a lower intensity.  {In addition, since we use a logarithmic scaling for the $y$ axis, we can only show the positive non-detection. } This explains why certain stacked trends fall below the distribution of significantly detected points in \autoref{fig:rgal_ratio} and \autoref{fig:rgal_ratio_2}.}. The observed decrease at 3\,kpc of individual sightlines is likely related to incomplete radial sampling.

We observe a slight positive correlation of $R_{21}^{12}$ with SFR surface density. Fitting a linear regression to the stacked ratio trend in log-log space, we find a slope of $m=0.08$. We note that this value is less steep, but within the expected range of scatter, than the slope of $m=0.129$ reported by \citet{Leroy2022} across a sample of 90 nearby galaxies at kpc-scale and $m=0.10\pm0.02$ measured by \citet{denBrok2023b} at ${\sim}200\,$pc scale resolution in NGC\,3627. Finally, we find a negative, statistically significant anticorrelation between the $^{12}$CO(1-0) FWHM and $R_{21}^{12}$.

\begin{deluxetable*}{l c |c c c| c c c}
\tablewidth{0pt}
\tablecaption{ Summary of Line Ratio Results \label{table:results_ratio}}
\tablehead{
\colhead{Ratio} & \colhead{$n_{\rm det}$\tablenotemark{a}}& \multicolumn{3}{c}{Weighted Mean and Scatter\tablenotemark{b}} & \multicolumn{3}{c}{Stacked Trends\tablenotemark{c} (slope/intercept/$p$-value)}  \\ && \colhead{entire map}  & \colhead{center} & \colhead{disk} & \colhead{$r_{\rm gal}$} & \colhead{$\Sigma_{\rm SFR}$} & \colhead{$\rm FWHM_{^{12}\rm CO(1-0)}$}}
\startdata
$R_{21}^{12}$ &2485& $0.78^{+0.14}_{-0.13}$ &$0.89^{+0.15}_{-0.15}$ &$0.76^{+0.14}_{-0.13}$ & ${-0}.031/{-0}.022/0.092$ & $0.08/{-0}.10/0.095$ & ${-0}.24/0.25/6.2\times 10^{-6}$ \\[5pt]
$R_{21}^{13}$ &298& $0.76^{+0.14}_{-0.17}$ &$0.87^{+0.29}_{-0.26}$ &$0.73^{+0.15}_{-0.16}$ & ${-0}.071/{-0}.072/0.018$ & ${0}.065/{-0}.14/0.03$ & $0.0089/{-0}.22/0.89$ \\[5pt]
$R_{21}^{18}$\tablenotemark{c} &65& $1.42^{+0.64}_{-0.76}$ &$1.57^{+0.47}_{-0.63}$ &$1.4^{+0.72}_{-0.83}$ & -- & -- & -- \\[5pt] \hline
$R_{10}^{13/12}$ &2001& $0.12^{+0.03}_{-0.03}$ &$0.15^{+0.03}_{-0.03}$ &$0.12^{+0.025}_{-0.027}$ & $-0.016/-0.9/0.56$ & $0.07/-0.82/2.1\times 10^{-5}$ & $-0.26/-0.55/3.7\times 10^{-6}$ \\[5pt]
$R_{21}^{13/12}$ &301& $0.13^{+0.02}_{-0.03}$ &$0.15^{+0.03}_{-0.03}$ &$0.12^{+0.021}_{-0.025}$ & $-0.057/-0.95/0.014$ & $0.10/{-}0.09/0.01$ & $-0.0058/-1/0.94$ \\[5pt]
$R_{10}^{18/13}$ &857& $0.24^{+0.04}_{-0.04}$ &$0.25^{+0.04}_{-0.04}$ &$0.23^{+0.039}_{-0.046}$ & $-0.07/-0.59/0.021$ & $0.038/-0.62/0.045$ & $0.014/-0.7/0.84$ \\[5pt]
$R_{10}^{18/12}$ &857& $0.03^{+0.01}_{-0.01}$ &$0.04^{+0.01}_{-0.01}$ &$0.029^{+0.0085}_{-0.0093}$ & $-0.1/-1.5/0.085$ & $0.081/-1.5/0.0049$ & $-0.24/{-1}.3/0.02$ \\[5pt]
\enddata

\tablecomments{ $^{\rm a}$ {Number of significantly detected sightlines.} $^{\rm b}$The mean weighted by the $^{12}$CO(1-0) integrated intensity. We also provide the weighted 16$^{\rm th}$-to-84$^{\rm th}$ percentile range. $^{\rm c}$We compute the slope, intercept, and Pearson $p$-value of the linear regression fit  to the stacked trend in log-log space. $^{\rm c}$The high line ratio averages are most likely reflecting a significant observing bias due to the low sensitivity of the C$^{18}$O(2-1) observations, for which low ratios are censored. $^{\rm d}$Discussed and presented in more detail in Gali\'c et al. (in prep.). }
\end{deluxetable*}

\subsubsection{\texorpdfstring{$R^{13}_{21}$}{Lg} Ratio}
Due to their fainter emission compared to the $^{12}$CO low-$J$ transition, we have fewer significantly detected sightlines for which we can report a significant $R^{13}_{21}$ ratio value. We report a $^{12}$CO(1-0) weighted average and 16$^{\rm th}$-to-84$^{\rm th}$ percentile range for the significantly detected sightlines of $\langle R_{21}^{13}\rangle^{\rm weight.}=0.76^{+0.14}_{-0.17}$. The stacked trend is within the margin of scatter of the galaxy-wide average of $\langle R_{21}^{13}\rangle^{\rm weight.}_{\rm lowres}=0.61^{+0.09}
_{-0.08}$ noted by \citetalias{denBrok2022} at $33''$ resolution. Similar to $R_{21}^{12}$, we measure a mild decreasing trend of the line ratio with increasing galactocentric radius. From the stacked line ratio trend, we measure a decreasing slope of $m=-0.07$. From $r_{\rm gal}=0.0-0.5$\, kpc, the decrease is stronger with a negative slope of ${\sim}-1.5$. Also similar to $R_{21}^{12}$, we find an increasing trend with the SFR surface density. With $m=0.065$ ($p$-value of 0.03) the slope in log-log space is slightly more shallow than the $R_{21}^{12}$ correlation in its value.  Finally, $R_{21}^{13}$ does not show a significant correlation with the FWHM of $^{12}$CO(1-0). We measure $m{\approx}0$ ($p$-value of 0.79) in log-log space for the stacked trend. We cannot rule out that the lack of any correlation is driven by the low S/N of the $^{13}$CO(2-1) line, which results in lower stacked intensities as we sum over a more significant fraction of noise. 

\subsubsection{\texorpdfstring{$R^{18}_{21}$}{Lg} Ratio}
For C$^{18}$O(2-1) we have the smallest number of significantly detected sightlines of only 65 sightlines where $R_{21}^{18}$ has $\rm S/N{\ge}5$ (compared to 2485 sightlines for $R_{21}^{12}$ and 298 sightlines for $R_{21}^{13}$; see percentage completeness in \autoref{tab:obs_lines}). For these significantly measured line ratios, we find a luminosity weighted average and 16$^{\rm th}$-to-84$^{\rm th}$ percentile range of $\langle R_{21}^{18}\rangle^{\rm weight.}=1.4^{+0.5}_{-0.6}$. We note that this is significant higher than the transition ratios of $^{12}$CO and $^{13}$CO we report here in this study. Furthermore, the CLAWS survey reported ratio at 33$''$ is also lower with $\langle R_{21}^{18}\rangle^{\rm weight.}_{\rm lowres}=0.87^{+0.24}_{-0.15}$ \citepalias{denBrok2022}, although in agreement within the scatter. In contrast, the stacked ratio is around 0.7, which is more comparable to the other CO line transition ratios. The reason for this high ratio measured for the individual sightline most likely is due to an observational bias. The C$^{18}$O(2-1) emission is only marginally detected across our map. Therefore, low $R_{21}^{18}$ line ratio values, where the 2-1 emission tends to be fainter, are likely censored and not significantly detected. Only when stacking can we probe the full parameter space and measure  lower $R_{21}^{18}$. Since only 2-3 stacked measurements are significant, we do not analyse  {or quantify} any trends of the line ratio present with galacocentric radius, SFR surface density, nor $^{12}$CO(1-0) line FWHM. 


\subsubsection{\texorpdfstring{$R^{13/12}_{10}$}{Lg} Ratio}
For $R_{10}^{13/12}$, we measure a weighted average and 16$^{\rm th}$-to-84$^{\rm th}$ percentile range of $\langle R_{10}^{13/12}\rangle^{\rm weight.}=0.12^{+0.03}_{-0.03}$. This is in agreement with the distribution of values of $\langle R_{10}^{13/12}\rangle^{\rm weight.}_{\rm lowres}=0.12^{+0.02}_{-0.02}$ found for M51 at low angular resolution \citepalias{denBrok2022}. At 33$''$ resolution, the ratio shows a mild decreasing trend with radius. With our high angular resolution observations, we see the radial line ratio trend has three distinct regimes: the ratio sharply increases with radius out to 0.6\,kpc. Such a trend has been reported by the previous studies by \citet{Tosaki2002}, although their ratio is a factor 2 lower (they report $R_{10}^{13/12}\sim$0.05--0.1). For radii ${\ge}0.6$\,kpc, the ratio then decreases as function of radius. From $0-0.6$\,kpc, the slope of the trend amounts to $m=0.25$. At ${\ge}0.6$\,kpc, the slope then changes to $m=-0.04$. Finally, at $r{>}2\,$kpc, where the transition from the central region to the onset of the spiral arms occurs, the ratio again mildly increases. 

We also detect an increasing trend of $R_{10}^{13/12}$ with SFR surface density. A linear regression of the stacked trend in log-log space (orange line) yields a slope of $m=0.07$ ($p$-value of $2.1\times10^{-5}$), which is comparable to the slope of the correlation of $R^{13}_{21}$ with the SFR surface density. Finally, the stacked line ratio trend does  show a significant negative correlation with the $^{12}$CO(1-0) line FWHM. We find a decreasing slope of $m=-0.27$ (the $p$-value ${\ll}0.01$) in log-log space.

\subsubsection{\texorpdfstring{$R^{13/12}_{21}$}{Lg} Ratio}
For $R^{13/12}_{21}$, we measure higher values than reported by \citetalias{denBrok2022} at coarse angular resolution. The luminosity weighted average and 16$^{\rm th}$-to-84$^{\rm th}$ percentile range at $4''$ resolution is $\langle R_{21}^{13/12}\rangle^{\rm weight.}=0.13^{+0.03}_{-0.03}$. For comparison, the distribution reported at $33''$  is  
$\langle R_{21}^{13/12}\rangle^{\rm weight.}_{\rm lowres}=0.09^{+0.01}_{-0.02}$.  The ratio shows a mild decrease with increasing radius (slope of $m=-0.06$ with $p$-value of 0.01). We do detect a mildly increasing ratio trend between $0-0.6$\,kpc, similar to the $R_{10}^{13/12}$ trend. The {stacked} ratio  {trend} correlates positively with the SFR surface density. In log-log space, we measure for the stacked line ratio trend a slope of $m=0.1$. This correlation is comparable to the other line ratio trends we measure.  Finally, $R^{13/12}_{21}$ does not show any correlation with the FWHM of $^{12}$CO(1-0) nor  $^{12}$CO(2-1) (not shown here), neither its stacked trend nor the individual significantly detected sightlines. 
 
\subsubsection{\texorpdfstring{$R^{18/13}_{10}$}{Lg} Ratio}
Because the emission of both line transitions of $R^{18/13}_{10}$ is optically thin, this ratio makes it possible to directly trace any change of the relative abundances of the $^{13}$CO and C$^{18}$O species. This particular ratio is studied and discussed in detail in Gali\'c et al. (in prep.) and we refer the reader to this publication for more details. 
Qualitatively, this ratio shows, similarly to $R_{10}^{13/12}$, a decreasing trend towards the center at $r_{\rm gal}{<}1\,$kpc. The ratio also decreases then again with increasing radius.

\subsection{Molecular Gas Conditions under LTE-Assumption} \label{sec:ltemodeling}
Assuming local thermal equilibrium (LTE), optically thick $^{12}$CO, and optically thin $^{13}$CO emission, we can use analytic expressions to derive the excitation temperature from $^{12}$CO and the optical depth and column density using $^{13}$CO. We choose $^{13}$CO over the other optically thin tracer C$^{18}$O because of the higher respective S/N of the line. 
In summary, we use the following specific assumptions \citep{Wilson2009}:
\begin{itemize}
    \item For a given sightline, the excitation temperature, $T_{\rm ex}$, is uniform for all rotational $J$ transitions.
    \item The excitation temperature is the same for all CO isotopologues.
    \item The emission of both $^{12}$CO(1-0) and (2-1) is optically thick ($\tau\gg1$).
    \item The $^{12}$CO and $^{13}$CO line emission originates from the same volume, as described by an identical beam filling factor, $\phi$.
\end{itemize}
These assumptions do not necessarily represent the real conditions in the ISM but can provide a quick and straightforward constraint on the gas density, temperature, species abundances and line optical depths (see \autoref{sec:radexmodeling} for our non-LTE calculations). The LTE approach is, often performed in the literature in the Milky Way and nearby galaxies \citep[e.g.,][]{Dickman1978,Nishimura2015,Cormier2018,Roueff2021, Wange2023}. Therefore, we will first focus on these assumptions, and provide a comparison to the results of other studies. 



\begin{figure*}
    \centering
    \includegraphics[width=\textwidth]{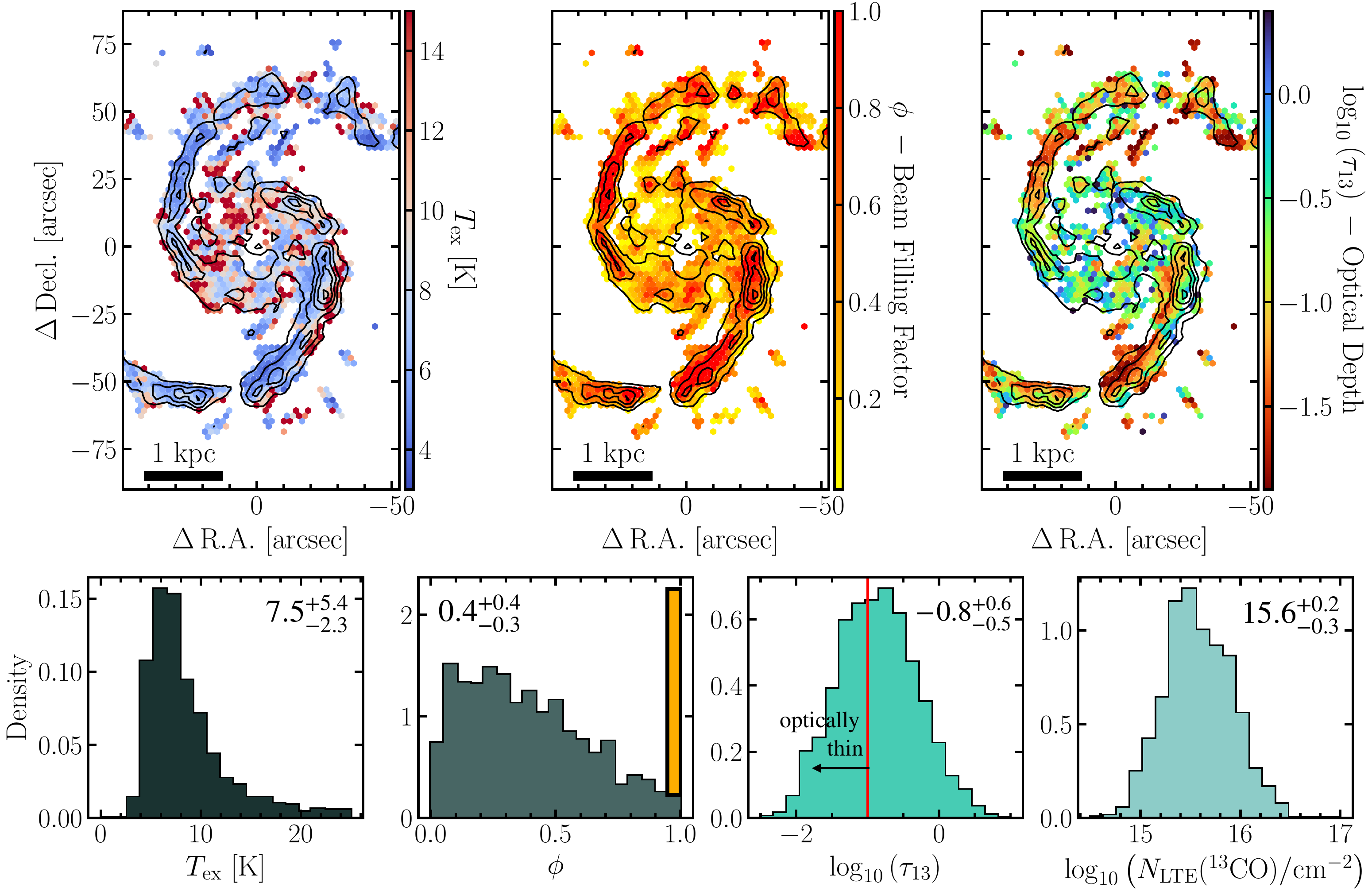}
    \caption{{\bf Distribution of LTE-derived parameter values} (\textit{Top row}) The maps illustrate the distribution of the excitation temperature, $T_{\rm ex}$ (left), the beam filling factor, $\phi$ (center), and the $^{13}$CO optical depth (right), across  {the inner region of} M51. The contours indicate the $^{12}$CO(1-0) integrated intensities ranging from 40$\rm~K\,km\,s^{-1}$ to 200$\rm~K\,km\,s^{-1}$ in steps of 40$\rm~K\,km\,s^{-1}$. (\textit{Bottom row}) The panels present the distribution of the excitation temperature ($T_{\rm ex}$), the beam filling factor ($\phi$), the optical depth of $^{13}$CO ($\tau_{13}$), and the $^{13}$CO column density ($N_{\rm LTE}(^{13}\rm CO)$). The median and 16$^{\rm th}$-to-84$^{\rm th}$ percentile range are indicated in the upper corner of each panel. The beam filling factor is set to $1$ if the derived value exceeds unity. This causes a pile-up (indicated in orange) in the distribution. Optical depth values below log$_{10}$$(\tau){<}-1$ (left of the red vertical line) can be considered as optically thin.}
    \label{fig:LTE_model1}
\end{figure*}

We start with the following equation 15.29 from \citet{Wilson2009}, which relates the peak temperature, $T_{\rm peak}$, of a line and the excitation temperature, $T_{\rm ex}$:
\begin{equation}\label{eq:tex}
T_{\rm peak} = \phi\left(1-e^{-\tau}\right)\frac{h\nu}{k_{\rm B}}\left[\frac{1}{e^{h\nu/k_{\rm B}T_{\rm ex}}-1}-\frac{1}{e^{h\nu/k_{\rm B}T_{\rm CMB}}-1}\right],
\end{equation}
where  {$\phi$ is respective the beam filling factor}, $\tau$ is the optical depth, $h$ the Planck constant, $k_{\rm B}$ the Boltzmann constant, $\nu$ the  restframe frequency, and $T_{\rm CMB}$ the cosmic microwave background temperature of $2.71\,$K. 
Then, under the set of assumptions outlined above, we proceed as follows:
\begin{itemize}
    \item We apply \autoref{eq:tex} to our $^{12}$CO(1-0), $^{12}$CO(2-1) and $^{13}$CO(1-0) line measurements. This yields three equations and three measurements (the peak brightness temperatures) for each line of sight.
    \item From this system of equations, we still remain with four unknowns: a common filling factor, a common $T_{\rm ex}$, and the $^{12}$CO(1-0) and $^{13}$CO(1-0) opacities\footnote{ {We note that, technically, we have another unknown parameter: the opacity of $^{12}$CO(2-1). However, for the purposes of these calculations, we only require $^{12}$CO(1-0) and $^{12}$CO(2-1) to be optically thick and do not derive any estimates of the optical depth for $^{12}$CO(2-1) itself.}}. 
    \item However, under the assumption of optically thick $^{12}$CO(1-0) and (2-1) emission, the term $e^{-\tau}$ will tend to 0, reducing the system to three unknown parameters.
    \item  We infer $T_{\rm ex}$ and the filling factor from the two $^{12}$CO equations and use them to determine the $^{13}$CO opacity. 
    \item {\bf additionally:} Under reasonable assumptions of the [$^{12}$CO]/[$^{13}$CO] abundance ratio, we can crosscheck that the assumption of $^{12}$CO opacity ${\gg}1$ is justified.  
\end{itemize}


The optical depth of $^{13}$CO is then computed using the following equation:
\begin{equation}\label{eq:od}
    \small
    \tau_{^{13}\rm CO} = -\ln\left[1-\frac{T_{\rm peak}^{13}}{\phi h\nu/k_{\rm B}}\left\{ \frac{1}{e^{\frac{h\nu}{k_{\rm B}T_{\rm ex}}}-1}-\frac{1}{e^{\frac{h\nu}{k_{\rm B}T_{\rm CMB}}}-1}\right\}^{-1}\right]
\end{equation}
Using this estimate of the optical depth, $\tau$, for $^{13}$CO, we can further calculate the entire column density of all level populations of $^{13}$CO along each sightline. The (beam averaged) column density of $^{13}$CO  relates to the excitation temperature, the $^{13}$CO(1-0) line center optical depth, and the integrated $^{13}$CO(1-0) intensity, $W_{\rm ^{13}CO(1-0)}$ as follows \citep[][]{Wilson2009}:
\begin{equation}\label{eq:cdens}
\small
\begin{split}
    N_{\rm LTE}(\rm ^{13}CO) \left[\rm {cm^{-2}}\right] =&\\ 3.0\times10^{14}\cdot \frac{1}{1-e^{-5.3/T_{\rm ex}}}\cdot& \frac{\tau_{^{13}\rm CO}}{1-e^{-\tau_{^{13}\rm CO}}} \cdot W_{\rm ^{13}CO(1-0)}
\end{split}
\end{equation}

\begin{figure*}
    \centering
    \includegraphics[width=0.96\textwidth]{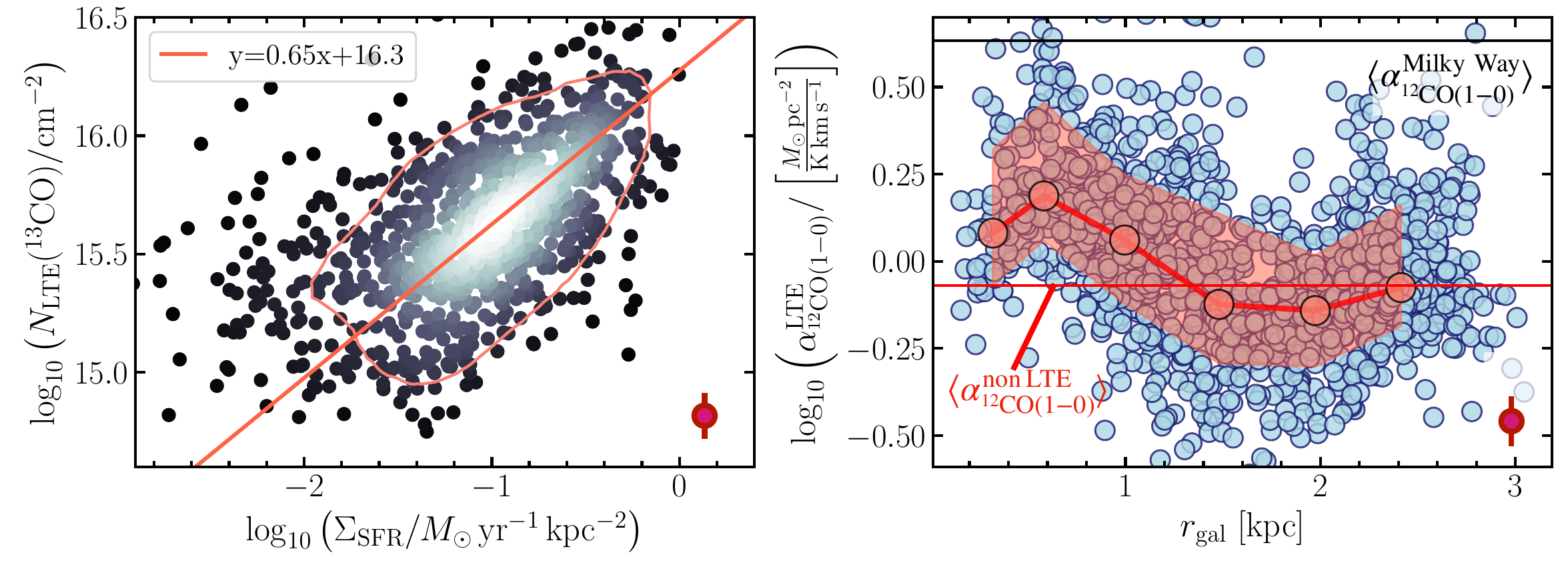}
    \caption{{\bf Correlations of the LTE-based parameters.} 
    (\textit{Left}) We compute the $^{13}$CO column density, $N_{\rm LTE}\left(^{13}\rm CO\right)$ based on the $^{13}$CO(1-0) integrated intensity using \autoref{eq:cdens}, which assumes LTE. Only sightlines were used where the $^{13}$CO(1-0) line emission is significantly detected ($\rm S/N\ge5$). 
    (\textit{Right}) The CO-to-H$_2$ conversion factor for the $^{12}$CO(1-0) emission as a function of galactocentric radius assuming [$^{13}$CO/H$_2$]$=1.7\times 10^{-6}$. The red circles show binned values. The shaded red area represents the $\pm1\sigma$ scatter per bin. For reference, we also present the local neighborhood value of 4.3 (black line) and the average value derived from our non-LTE analysis (red line). In the bottom right corner of each panel, we present the average error for a single datapoint, based on the MC resampling approach. In the left panel points are color-coded based on their density in the panel (this is a qualitative representation of how densely the points are distributed in the parameter space).}
    \label{fig:LTE_model2}
\end{figure*}

We present the distribution of values from our LTE-based calculations in \autoref{fig:LTE_model1} and investigate trends in \autoref{fig:LTE_model2}. We additionally assess the measurement uncertainty associated with individual data points, which we illustrate in the bottom right corner of each panel of \autoref{fig:LTE_model2}. This is achieved through a Monte Carlo (MC) resampling procedure conducted 100 times, wherein we systematically introduce the measured intensity's uncertainty and reiterate the LTE-modeling process. Subsequently, we quantify the uncertainty by calculating the standard deviation across the 100 iterations for all modelled quantities.


\subsubsection{Excitation Temperature in M51}
We present the derived excitation temperature in the top and bottom left panels of \autoref{fig:LTE_model1}. The map showcases the spatial distribution of $T_{\rm ex}$. Qualitatively, we do not find any clear spatial trends apart from a slight increase  toward the downstream region of the southern spiral arm, where heating from young stars could play a role, and in the central northern region the values also are higher than in the spiral arm. The bottom left panel of \autoref{fig:LTE_model1} presents the distribution of the values across  {the mapped region of} M51. We measure a mean and 16$^{\rm th}$-to-84$^{\rm th}$ percentile range of $\langle T_{\rm ex}\rangle = 7.5^{+5.4}_{-2.3}\,\rm K$. While most temperatures have $T_{\rm ex}<10\,\rm K$ (80\% of the sightlines), there is a small tail towards higher excitation temperatures exceeding 10\,K, indicating regions of more extreme excitation conditions. We note again that sightlines where the $^{12}$CO(2-1)/(1-0) peak temperature ratio exceeds unity (2\% of sightlines) are excluded from the subsequent analysis. Such sightlines require accounting for non-LTE effects, which is part of the analysis presented in \autoref{sec:radexmodeling}. 

\subsubsection{The beam filling factor for \texorpdfstring{$^{12}$CO}{Lg}}

The top middle and bottom second to left panels in \autoref{fig:LTE_model1} present the distribution of derived $\phi$ values. We note that 5\% of the sightlines show a beam filling factor larger than unity. This likely indicates that, in general, one or more of the assumptions listed at the beginning of this subsection, are not fully met for these sightlines. Assessing qualitatively the top middle panel in \autoref{fig:LTE_model1}, we see that the beam filling factor reaches values close to unity along the central spine of the spiral arm and decreases towards its edges. In the interarm region, the beam filling factor is consistently below unity. Overall, we find a peak in the $\phi$ distribution between $0.1$ and $0.3$, and a  piling up of values at 1  (which is artificially induced, as we cap the beam filling factor at unity). 

\subsubsection{The optical depth of \texorpdfstring{$^{13}$CO}{Lg}}
In the subsequent analysis, we focus on the optical depth derived using the $^{13}$CO(1-0) emission line according to \autoref{eq:cdens}. 
The overall distribution of $\tau_{\rm ^{13}CO}$ is represented in the bottom middle panel in \autoref{fig:LTE_model1}. We note that apart from a small fraction ($6\%$ in area) it is less than unity everywhere across the galaxy at $4''$ (${\sim}170$\,pc) scales. We find a mean of $\langle\log_{10}\left(\tau_{^{13}\rm CO}\right)\rangle=-0.8^{+0.6}_{-0.5}$. 44\% sightlines have an optical depth of $\log_{10}\left(\tau_{^{13}\rm CO}\right){<}{-1}$, which we consider as threshold for optically thin emission. A large part of the optical depth therefore are neither optically thin nor optically thick, but lie in the regime between. These elevated opacities suggest potential limitations of $^{13}$CO as a tracer of the entire column of gas along the line of sight.  As a consistency check, assuming a [$^{12}$CO]/[$^{13}$CO] abundance ratio of 30 for M51 \citep[e.g.,][which is consistent with the value we find using the non-LTE approach, see Section \ref{sec:radexmodeling}]{Schinnerer2013}, 67\% of sightlines have an optical depth $\tau{>}2$, implying optically thick $^{12}$CO emission.


\subsubsection{The \texorpdfstring{$^{13}$CO}{Lg} Column Density and  \texorpdfstring{$\alpha_{\rm CO}$}{Lg}}
\label{sec:alpha_CO_nonLTE}
In the left panel of \autoref{fig:LTE_model2} we present the LTE-derived $^{13}$CO column densities using \autoref{eq:cdens} as a function of SFR surface density (based on the 33\,GHz continuum emission). The two quantities appear to correlate (Pearson's $p$ value ${\ll}0.05$). This is particularly evident if we fit a linear regression to the sightlines within the 80\% inclusion region (indicated by the red contour in the panel). The observed positive correlation between column density and SFR surface density is in line with the Schmidt-Kennicutt relation \citep{Schmidt1959,Kennicutt2012}, which connects the surface density of (molecular) gas with the SFR. We note, however, that the slop is shallower than the one found for $\Sigma_{\rm SFR}$ and $\Sigma_{\rm mol}$ in M51 \citep[][report a slope of ${\sim}1.2$]{Bigiel2008}. Directly comparing the Schmidt-Kennicutt relation is challenging as the  {derived} slope is  {affected when applying} different sensitivities and depends on the treatment of upper limit.
 { Therefore, although our approach estimates the column density differently -- using LTE modeling rather than applying a constant CO-to-H$_2$ conversion factor to CO intensities -- as commonly done in the literature, the  correlation between the $^{13}$CO column density and the SFR surface density remains robust. Therefore, this suggests that the column density derived from our method still effectively traces the mass distribution of molecular gas.}



Therefore, using a fiducial $^{13}$CO-to-H$_2$ abundance ratio, we can also translate the $^{13}$CO column density into an estimate for the H$_2$ column density. More generally, we are interested in establishing an estimate for the CO-to-H$_2$ conversion factor ($\alpha_{\rm ^{12}CO(1-0)}$, hereafter just abbreviated as $\alpha_{\rm CO}$ ), which is defined as the ratio of molecular gas surface density to the integrated $^{12}$CO(1-0) intensity \citep{Bolatto2013}. Following \citet{Kamenetzky2014} and \citet{Teng2022}, we can derive the following functional form for the conversion factor as function of the $^{13}$CO column density:
\begin{equation}\label{eq:aco_LTE}
    \alpha^{\rm LTE}_{\rm ^{12}CO(1-0)}=\frac{1}{4.5\times 10^{19}}\frac{N(^{13}\rm CO)/\left[\rm ^{13}CO/H_2\right]}{W_{^{12}\rm CO(1-0)}}
\end{equation}
The derivation of the conversion factor depends on the assumed abundance ratio. For this analysis, we assume an abundance of $\left[^{13}\rm CO/\rm H_2\right]=1.7\times 10^{-6}$ \citep{Dickman1978}. We note, however, that the abundance ratio varies within galaxies \citep[e.g.,][]{Frerking1982, Goldsmith2008, Sliwa2012} by up to an order of magnitude with a typical scatter around 0.5\,dex for a fixed H$_2$ density \citep{vanDishoek1992,Sheffer2008}. At cloud-scales, we do not resolve the molecular clouds, therefore, the beam-to-beam variations of the abundance are expected to be smaller. Furthermore, within the central $r{\le}3\,\rm kpc$ region which we analyze here, we do not expect such large variations, as large variations in the $^{13}$CO abundance are mostly related to lower metallicity and lower (column) density regimes, such as the outskirts of galaxies. 

The right panel of \autoref{fig:LTE_model2} illustrates the derived conversion factor as a function of galactocentric radius. We find a mean conversion factor distribution  {and a corresponding $1\sigma$ scatter per bin/region} of $\langle \alpha^{\rm LTE}_{\rm ^{12}CO(1-0)}\rangle=1.0^{+0.2}_{-0.2}\,M_\odot\,\rm pc^{-2}/(K\,km\,s^{-1})$. The values towards the center ($r_{\rm gal}<0.2$\,kpc) are significantly lower with $\langle \alpha^{\rm LTE}_{\rm ^{12}CO(1-0),center}\rangle=0.48^{+0.03}_{-0.03}\,M_\odot\,\rm pc^{-2}/(K\,km\,s^{-1})$. The conversion factor increases with radius up to ${\sim}0.5\,$kpc. At larger radii, we measure a negative correlation with galactocentric radius.  We note two things in particular when comparing the derived conversion factor relation to the one described in \citet{denBrok2023} at coarser, kpc-scale, resolution: (i) our average CO-to-H$_2$ conversion factor is ${\sim}$3 times lower (they report a value of 3.1\,$M_\odot\,\rm pc^{-2}/(K\,km\,s^{-1})$ for the central 2kpc in radius), (ii) they do not report a lower conversion factor towards the center. In their study, \citet{denBrok2023} measured the conversion factor using a dust-based approach to estimate independently the H$_2$ mass distribution. Lower systematic $^{13}$CO-based conversion factors have been found before \citep[e.g.][]{Cormier2018}, who find, on average, a factor 2 lower than the Milky Way value for their sample of galaxies. It is worth noting that the conversion factors based on  $^{13}$CO, are predicted to be offset by a factor of 2-3 by simulations \citep[e.g.,][]{Szucs2016}, which would bring our values and the dust-based values closer in agreement. 
One explanation for this is that the excitation temperatures for the different species can differ. Looking at \autoref{eq:cdens}, the conversion factor scales approximately linearly with excitation temperature (for optically thin lines).  Finally, we also have not considered any gradient in the $^{13}$CO-to-H$_2$ abundance ratio, which could explain the discrepancy as well. A linear relation of the $^{12}$CO/H$_2$ abundance with metallicity is expected from theoretical considerations \citep{Bialy2015}, but over the small radial range we cover in M51, such a gradient is not significant. For instance,  \citealt{Berg2020} report a radial metallicity gradient of 0.2\,dex over the entire disc of M51. This suggests only gradients and changes in the relative $^{12}$CO/$^{13}$CO abundance ratio remain relevant. 
For a more robust approach, which takes differences in the excitation temperature and varying $^{12}$CO-to-$^{13}$CO abundance ratios into consideration we require non-LTE computations.

\subsection{Beyond LTE: Modeling the Line Emission by solving  the non-LTE Radiative Transfer Equations}
\label{sec:radexmodeling}

As an alternative approach to the LTE-based calculations, we employ the non-LTE radiative transfer code RADEX \citep{vanTak2007} to model the observed line intensities. Here, we follow the framework presented in \citet{Teng2022,Teng2023}, and refer the reader to these publications for details and background on the RADEX implementation. The free parameters that we model are the permutations of a range of kinetic temperature ($T_{\rm kin}$), H$_2$ volume density ($n_{\rm H_2}$), $^{12}$CO column density per line width ($N_{\rm CO}/\Delta v$), the  $^{12}$CO/$^{13}$CO abundance ratio ($\left[^{12}\rm CO/^{13}CO\right]$), and the beam filling factor ($\phi$). 
As input for the energy levels, statistical weights, Einstein $A$, and collisional rate coefficients of each CO isotopologue we use the data files from the Leiden Atomic and Molecular Database \citep[LAMDA;][]{Schoier2005} which provide collisional rate coefficients from \citet{Yang2010}. 
We note that \citet{Tunnard2016} have shown that such a model setup can recover the physical conditions of the molecular gas using CO isotopologues. For the non-LTE line modeling, we do not include  the C$^{18}$O (1-0) and (2-1) lines as the (2-1) observations have a low sensitivity and hence would limit the analysis to a very small number of sightlines.

\begin{figure*}
    \centering
    \includegraphics[width=\textwidth]{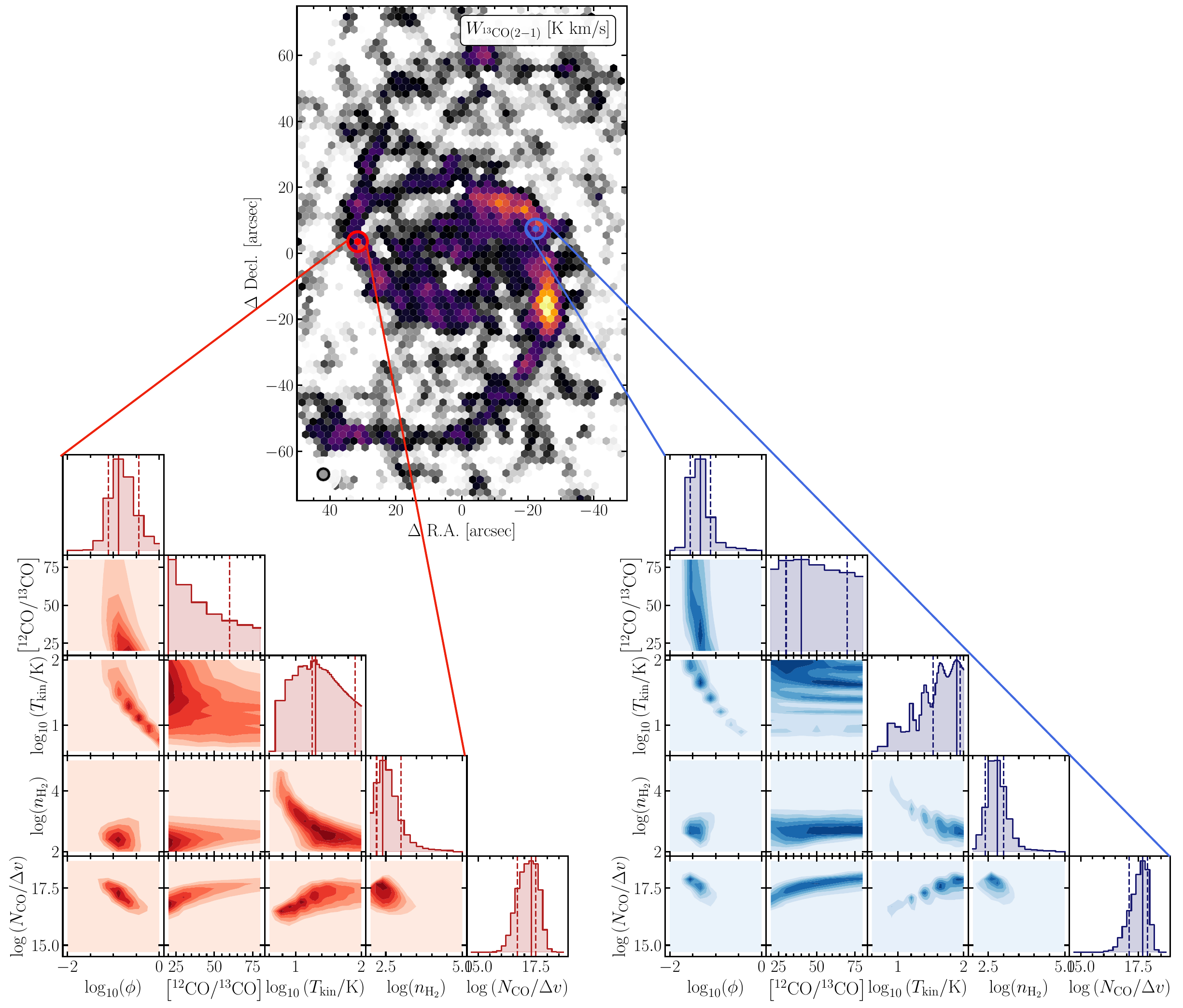}
    \caption{{\bf The 1D and 2D Marginalized Likelihood Distributions.} The red (left) and blue (right) corner plots show the resulting  1D and 2D likelihood probability density function from $\chi^2$ minimization following \autoref{eq:probdens}. The map in the center illustrates the $^{13}$CO(2-1) integrated intensity with the sightlines highlighted for which the two corner plots are calculated.  }
    \label{fig:zoomin_model}
\end{figure*}

\begin{figure*}
    \centering
    \includegraphics[width=0.9\textwidth]{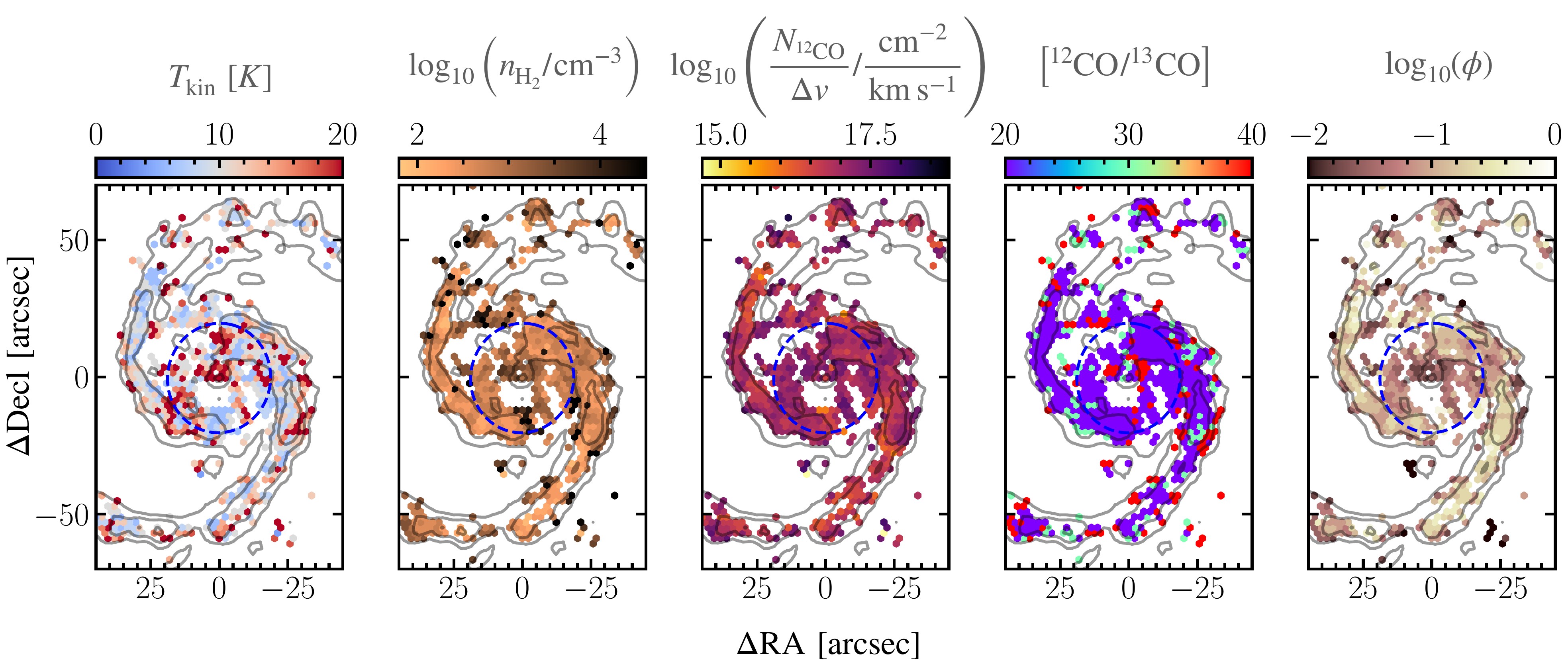}
    \caption{{\bf Spatial Distribution of the maximum likelihood model.} These panels show the pixel-wise maximum likelihood value derived from the $\chi^2$ minimization approach. (\textit{Left}) The kinetic temperature, $T_{\rm kin}$, map. The contours show the integrated $^{12}$CO(1-0) intensity isophote at 20, 50, 100, and 200 K\,km\,s$^{-1}$. The blue-dashed circle indicates 20$''$ in radius, which we use to differentiate between center and disk. (\textit{Second left}) The distribution of the mean volume density, $n_{\rm H_2}$, from which the CO emission is originating. (\textit{Middle}) The $^{12}$CO column density map. (\textit{Second right}) The beam filling factor. (\textit{Right}) The abundance ratio of $^{12}$CO-to-$^{13}$CO map. We note that we only have solutions for pixels where all four lines ($^{12}$CO and $^{13}$CO (1-0), (2-1)) are significantly (${>}5\sigma$) detected.}
    \label{fig:map_model}
\end{figure*}

\begin{figure*}
    \centering
    \includegraphics[width=0.8\textwidth]{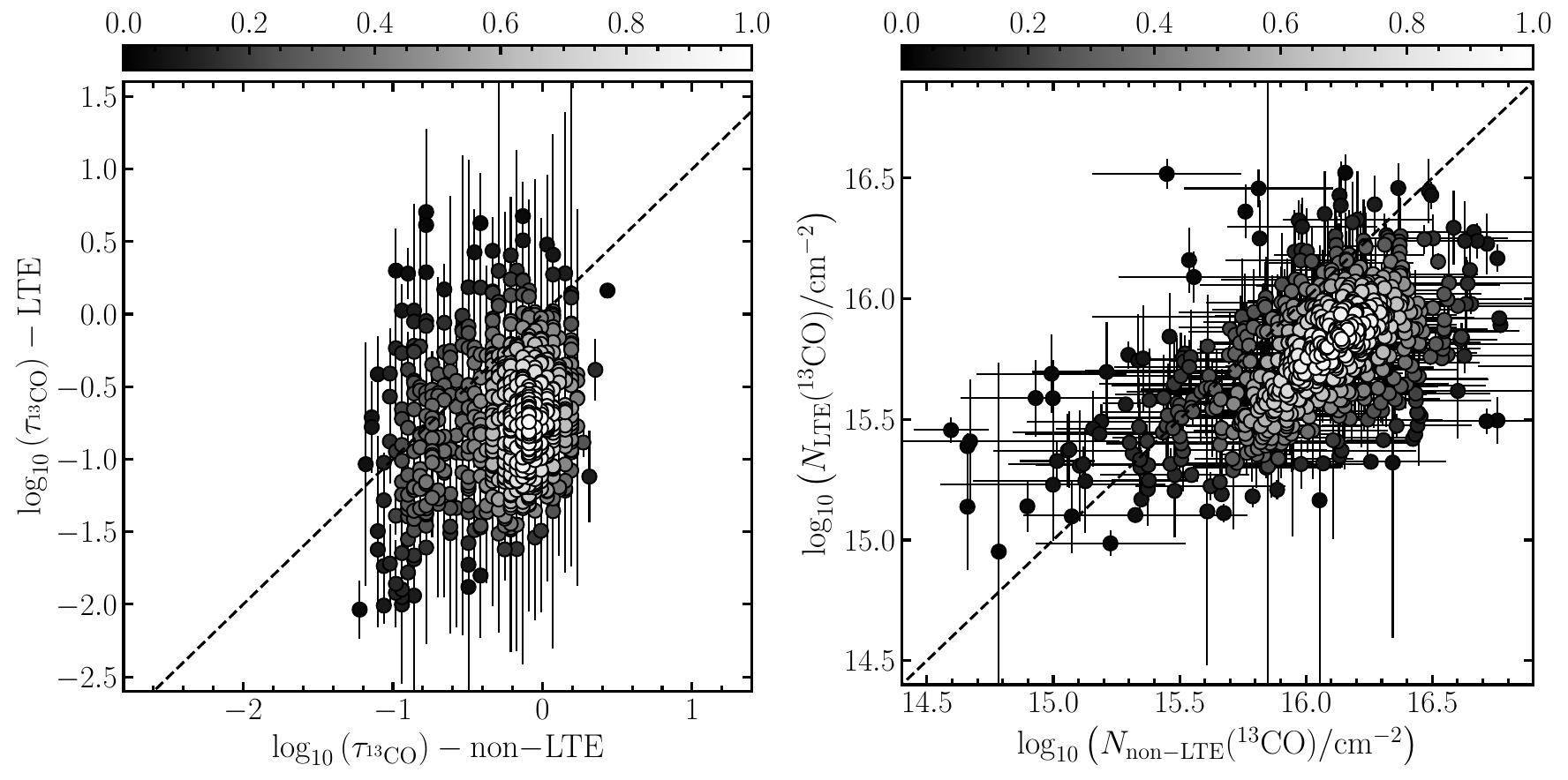}
    \caption{{\bf Comparing LTE and non-LTE derived parameters.} (\textit{Left}) Contrasting the optical depth of $^{13}$CO(1-0). The points are color-coded by their parameter-space filling based on a 2D Gaussian kernel function. The black dashed line illustrates the 1:1 relation. (\textit{Right}) Comparing the total $^{13}$CO column density derived using the LTE and non-LTE approach. The panel description follows the left one.  }
    \label{fig:comp_model}
\end{figure*}

We run \texttt{RADEX} by creating first a four-dimensional grid with the grid-points specified as $(T_{\rm kin}$, $n_{\rm H_{2}}$, $N_{\rm CO}/\Delta v$, $\left[^{12}\rm CO/^{13}CO\right]$).  {We then expand this grid to five dimensions by multiplying the intensities by a range of beam filling factors ($\phi$)}. For this model, we assume that these conditions are uniform for a given line of sight. The range of kinetic temperature, H$_2$ density, CO column density, and abundance ratio used to build the grid is presented in \autoref{tab:mode_param}.
Moreover {, for the $N_{\rm CO}/\Delta v$ parameter,} we use a fixed line width of $\Delta v{=}15$\,km\,s$^{-1}$. While we expect the line width to scale with the size of the GMCs \citep{Heyer2004}, at a resolution of ${\sim}120\,$pc, we can hardly resolve the clouds. The line width of ${\sim}15$\,km\,s$^{-1}$ is hence empirically motivated (see \autoref{fig:rgal_ratio})\footnote{ {In essence, we  fit $N_{\rm CO}/\Delta v$ with \texttt{RADEX}, instead of fitting $N_{\rm CO}$ and $\Delta v$ separately.  The actual estimation of $N_{\rm CO}$ hence varies with the line widths. Therefore, where we provide estimates of  $N_{\rm CO}$ values, we multiply $N_{\rm CO}/\Delta v$ with either ${\sim}15$\,km\,s$^{-1}$ for the modelled column density, or with the observed CO(1–0) line width for each pixel.}}. In total, this grid amounts to ${\sim}$260,000 grid points. 
We provide the entire machine-readable parameter grid in Appendix \ref{app:radex_grid}.

\begin{table}[h]
    \centering
    \caption{\texttt{RADEX}-model parameters}
    \label{tab:mode_param}
    \begin{tabular}{c c c} \hline
         Parameter & Range & Step Size  \\ \hline \hline 
          $T_{\rm kin}$& 4 -- 100 \,K & 2\,K  \\
          $\log_{10}\left(n_{\rm H_2}/\rm cm^{-3}\right)$ & 2 -- 5 & 0.2\,dex \\
          $\log_{10}\left(N_{\rm ^{12}CO}/\rm cm^{-2}\right)$ & 16 -- 20 & 0.2\,dex \\
          $\left[^{12}\rm CO/^{13}CO\right]$ & 20 -- 80 & 10 \\
          $\Delta v$ & 15 \,km\,s$^{-1}$ & -- \\\hline
          $\log_{10}(\phi)$ & ${-}2$ -- 0 & 0.2\,dex \\\hline
    \end{tabular}
\end{table}

We perform a $\chi^2$ minimization to determine which grid point best represents the molecular conditions that produce the observed CO isotopologue line emission per sightline. We compute $\chi^2$ for the grid point $\theta = (T_{\rm kin}, n_{\rm H_{2}}, N_{\rm CO}/\Delta v, \left[^{12}\rm CO/^{13}CO\right],\phi)$ for the $n=4$ lines (the two lowest rotational $J$ transitions for $^{12}$CO and $^{13}$CO) using the following expression:
\begin{equation}\label{eq:chi2}
    \chi^2(\theta) = \sum_{i=1}^n\left(\frac{ {W}_i^{\rm model}(\theta) -  c^{\rm line}_i\cdot  {W}_i^{\rm obs}}{\sigma_i^{\rm obs}}\right)^2
\end{equation}
where $c^{\rm line}_i$ captures the adjustment of the line width. We use $c^{\rm line}_i=15\,{\rm km\,s^{-1}}/{\rm FWHM_i}$.
In \autoref{eq:chi2}, $W_i^{\rm model}$ and  $W_i^{\rm obs}$ represent the modelled and observed integrated intensity of the $i$-th line respectively. We adjust the modeled intensity by the beam filling factor of the line, $\phi_i$, which we vary from 0.01 -- 1 in logarithmic steps of 0.2\,dex. We assume a {n identical} beam filling factor for both the $^{12}$CO and $^{13}$CO emission lines. The observational uncertainty is captured in the equation by the parameter $\sigma_i$. For the purpose of the $\chi^2$ minimization approach, we use additionally to the noise uncertainty a conservative estimate of 10\% uncertainty on the measured intensity, reflecting the calibration uncertainty, which is commonly adopted in the literature \citep[e.g.,][]{Leroy2017dens, Teng2022}. To quantify the significance of the minimum $\chi^2$, we compute for each grid point a likelihood probability, assuming a multivariate Gaussian probability distribution:
\begin{equation}\label{eq:probdens}
    P( {W}^{\rm obs}|\theta) = \left(\prod_i^{n}(2\pi \sigma_i^2)^{-\frac{1}{2}} \right) \cdot e^{-\frac{1}{2}\chi^2(\theta)}
\end{equation}

We attribute to each sightline the parameter combination based on the marginalized 1D likelihoods. For the \texttt{RADEX} grid parameters, which are sampled uniformly, the marginalized 1D likelihood distribution is obtained by summing over all likelihoods per fixed parameter.  To derive the marginalized likelihoods of the $^{12}$CO and $^{13}$CO optical depths and $\alpha_{\rm CO}$, which are functions of the intrinsic grid parameters and possibly not regularly sampled, we proceed slightly differently. We produced a probability-weighted histogram based on the likelihoods for the optical depths. To account for the irregular sampling, we normalize this histogram by the uniformly weighted histogram.  We note that this approach is consistent with the approach described in \citet{Teng2022} (section 4.4).

\autoref{fig:zoomin_model} illustrates the method for two arbitrarily selected pixels of the map. The corner plots depict the 1D and 2D probability density function distribution (PDF; \autoref{eq:probdens}). The 1D and 2D PDFs are derived from the  {5D} PDF (since we have five free parameters) by summing along the remaining axes. { {In essence, we use four observed emission lines to derive constraints on five free parameters. This is possible because the $\chi^2$ minimization approach quantifies the probability of every combination of model parameters given the data. Such an approach is is well established in the astronomy-applied statistics literature \citep[e.g.,][]{Ivezic2014}. The derived ``best-fit" parameters are then estimated by marginalizing over these likelihood distributions. This differs from the classical fitting of the data where one can, for instance, analytically derive a ``best-fit" parameter.}} 

\autoref{fig:map_model} illustrates the spatial distribution of the derived physical quantities.  \autoref{tab:nonLTE results} lists the intensity weighted averages and scatter. The 1D and 2D pdfs of these two selected pixels represent well the general behavior of the $\chi^2$ minimization approach. Overall, the beam filling factor ($\phi$), the volume density ($n_{\rm H_2}$), and $^{12}$CO column density 
($N_{\rm ^{12}\rm CO}$) are well constrained. The $^{12}$CO-to-$^{13}$CO abundance ratio and the kinetic temperature ($T_{\rm kin}$) on the other hand are less well constrained as indicated by a wider distribution of values.

Similarly to the LTE line modeling approach, we estimate the uncertainty of an individual data point by resampling the data by adding noise to the intensities and repeating the non-LTE $\chi^2$-minimization. The uncertainties are then determined for each quantity by taking the standard deviation along the 100 samples.

{\renewcommand{\arraystretch}{1.3} %
\begin{table}
    \centering
    \footnotesize
    \caption{{\bf Non-LTE-based intensity weighted averages }}
    \label{tab:nonLTE results}
    \begin{tabular}{lccc} \hline 
         & all & center & disk   \\ \hline \hline
         $\log_{10}\left(\tau_{^{12}\rm CO}\right)$&$0.9^{+0.2}_{-0.3}$&$1.0^{+0.2}_{-0.2}$&$0.9^{+0.3}_{-0.2}$\\
         $T_{\rm kin}$ [K] &$13^{+5}_{-5}$&$17^{+3}_{-9}$&$12^{+4}_{-4}$\\
         $\log_{10}\left(n_{\rm H_2}/\rm cm^{-3}\right)$&$2.8^{+0.6}_{-0.4}$&$2.9^{+0.3}_{-0.3}$&$2.7^{+0.7}_{-0.3}$\\
         $\log_{10}\left(N_{^{12}\rm CO}/\Delta\,v\,\rm \frac{\rm km}{\rm cm^{-2}\,km^{-1}\,s}\right)$&$16.9^{+0.4}_{-0.4}$&$17.0^{+0.3}_{-0.1}$&$16.9^{+0.4}_{-0.4}$\\
         $\left[^{12}\rm CO/^{13}CO\right]$&$25^{+15}_{-5}$&$23^{+3}_{-3}$&$25^{+15}_{-5}$\\
         $\phi$&$0.2^{+0.1}_{-0.2}$&$0.1^{+-0.1}_{-0.1}$&$0.2^{+0.2}_{-0.2}$\\
         $\alpha_{\rm ^{12}CO(1-0)}^{\rm non-LTE}\ \left[\frac{M_\odot\,\rm pc^{-2}}{\rm K\,km\,s^{-1}}\right]$&$2.4^{+0.7}_{-0.9}$&$2.7^{+0.8}_{-0.6}$&$2.3^{+0.9}_{-0.8}$\\
         \hline
    \multicolumn{4}{l}{%
    \begin{minipage}{9cm}%
     \small {\bf Note}: The value corresponds to the $ ^{12}$CO(1-0) intensity weighted average and 16$^{\rm th}$-to-84$^{\rm th}$ percentile range. We differentiate also between center ($r_{\rm gal}{\le}20''$) and disk ($r_{\rm gal}{>}20''$).%
  \end{minipage}%
}\\
    \end{tabular}
\end{table}
}

\subsubsection{Comparison between LTE and non-LTE derived properties}
Using the non-LTE approach, we can directly compare the derived optical depth and the total column density of $^{13}$CO to the LTE-based results (discussed in \autoref{sec:ltemodeling}). We present the comparison in \autoref{fig:comp_model}. Note that we scale the \texttt{RADEX}-derived column densities by the beam-filling factor, $\phi$, such that both the LTE and non-LTE represent the beam-averaged column density. Furthermore, we correct the non-LTE value by the full observed $^{12}$CO(1-0) line width instead of the fiducial 15\,km\,s$^{-1}$ line width.   

Overall, the optical depth of $^{13}$CO(1-0) (left panel)  show on average a  distribution of values that is offset by 0.6\,dex with respect to the LTE values, with non-LTE values having a mean of $-0.2$. For the total column density of $^{13}$CO (right panel), the distributions also differ, as the non-LTE values are systematically higher by ${\sim}0.2$\,dex. The correlation between LTE and non-LTE column densities suggests that to first order, the intensities are proportional to the column density, as expected for line that is not optically thick. In \autoref{tab:comp_LTE_nonLTE} we provide mean and 16$^{\rm th}$-to-84$^{th}$ percentile range of the overall distribution of values of the optical depth and column density.

For the optical depth, we find that the non-LTE-based values are also $\tau_{\rm ^{13}CO}^{\rm non-LTE}{<1}$, indicating that the$^{13}$CO line emission is not optically thick. However, the values are also $\tau_{\rm ^{13}CO}^{\rm non-LTE}{>0.1}$, indicating that the line neither is  optically thin. This likely also explains to some degree why the LTE-derived column densities are offset by a factor of 2 as the assumption of optically thin emission is not necessarily true.  The scatter of the overall distribution of values is significantly larger for the LTE-derived ones with ${\sim}0.6\,\rm dex$ compared to ${\sim}0.2\,\rm dex$ for the non-LTE derived values. The larger scatter for LTE is expected since this approach has a lower number of free parameters, thereby limiting the ways the parameters can balance each other out.    

The total $^{13}$CO column density values of the overall distribution are systematically offset beyond the scatter for the LTE and non-LTE approach. The LTE-based mean and percentile range of $\langle\log_{10}\left(N_{\rm LTE}(^{13}\rm CO)/\rm cm^{-2}\right)\rangle=15.8^{+0.4}_{-0.3}$ is smaller than the non-LTE based distribution of values of $\langle\log_{10}\left(N_{\rm non{-}LTE}(^{13}\rm CO)/\rm cm^{-2}\right)\rangle=16.0^{+0.2}_{-0.2}$. Using a linear regression fit in logarithmic space, we find a significant correlation with Pearson's $p\ll0.05$. 

The overall agreement or correlation between the LTE and non-LTE derived \emph{averages} of the column density suggests that any discrepancies of further parameters are likely due to differences in the initial assumptions (e.g., of a fixed $^{12}$CO or the $^{13}$CO-to-H$_2$ abundance ratio). 

\subsubsection{The \texorpdfstring{$^{12}\rm CO$}{Lg}(1-0) optical depth and the \texorpdfstring{$^{12}\rm CO$}{Lg}-to-H\texorpdfstring{$_2$}{Lg} Conversion Factor}

\begin{figure}
    \centering
    \includegraphics[width=0.95\columnwidth]{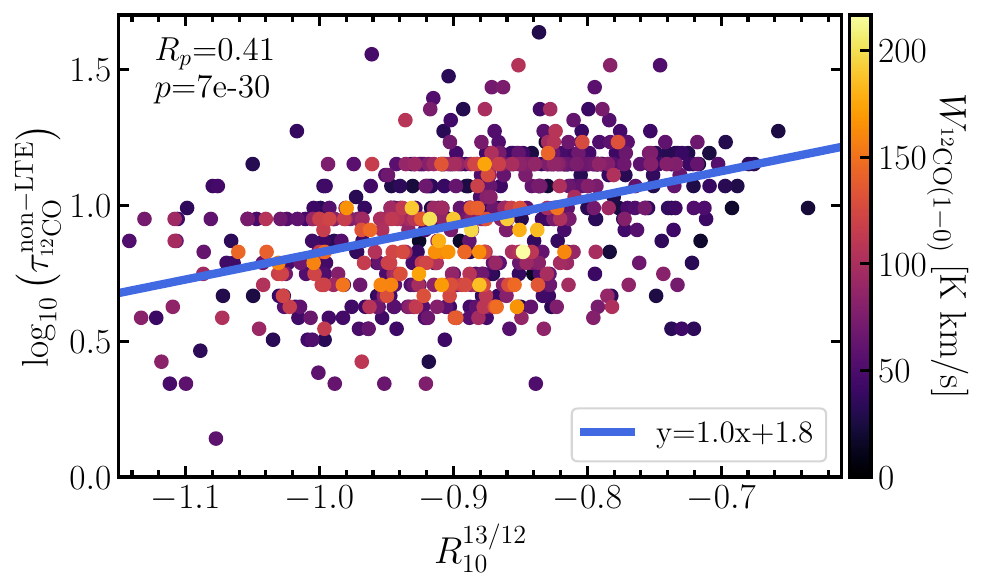}
    \caption{{\bf Optical depth of $^{12}$CO and the $R_{10}^{13/12}$ Line Ratio.} The points are color-coded with respect to the sightline's corresponding integrated $^{12}$CO(1-0) line brightness.  We perform a linear regression (in log-log space; blue line), which highlights the positive relation (Pearson's correlation coefficients indicated in top left corner) with the $R_{10}^{13/12}$ line ratio. }
    \label{fig:optical_depth}
\end{figure}

\begin{figure}
    \centering
    \includegraphics[width=0.9\columnwidth]{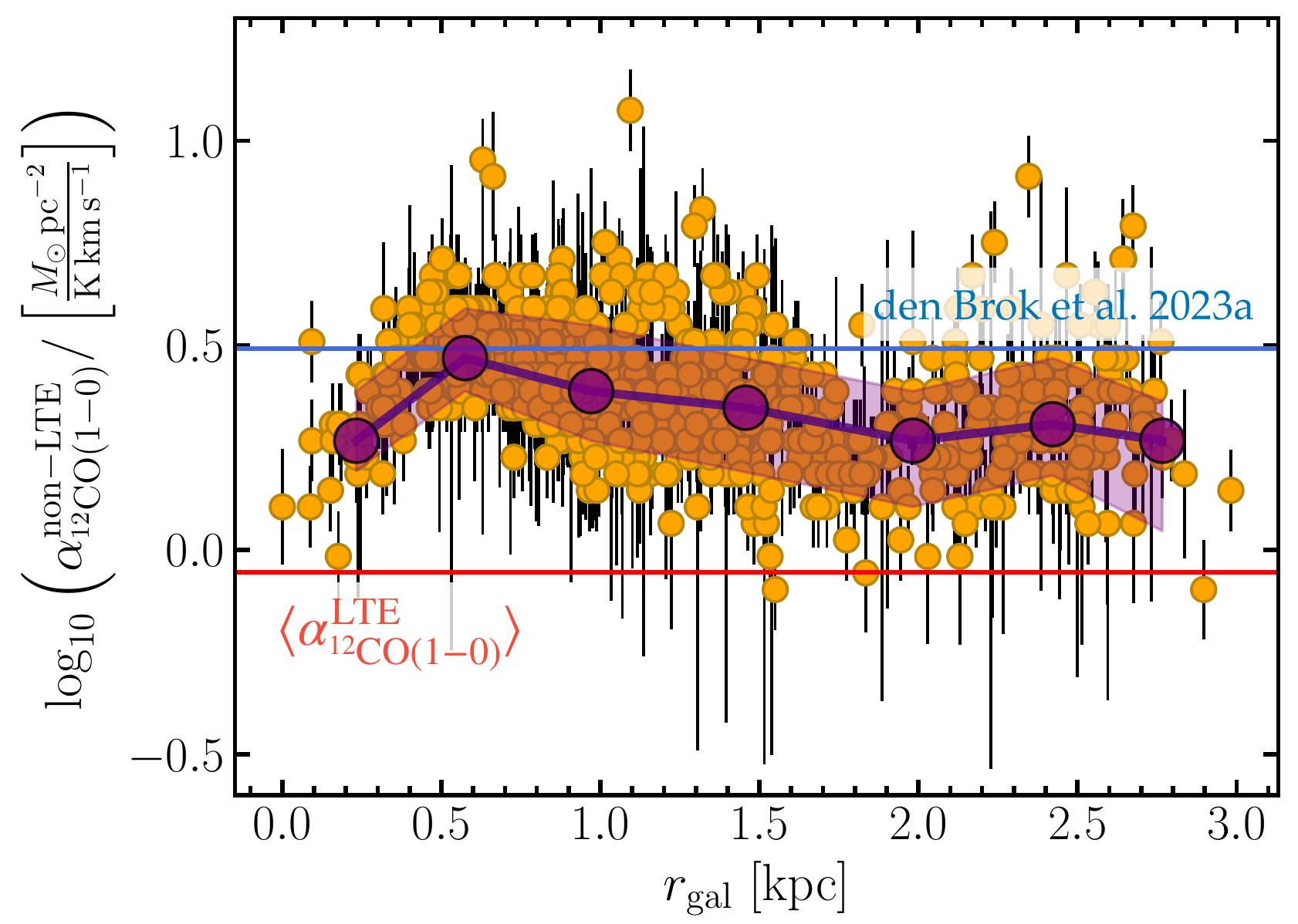}
    \caption{{\bf Radial Trend of the non-LTE derived CO-to-H$_2$ Conversion Factor.} We find overall a flat trend of $\alpha_{\rm ^{12}CO(1-0)}$ with galactocentric radius. The orange points show the value for the individual sightlines. The purple points depict the radially binned average. We illustrate the scatter of the radially binned values using the purple shade. The galaxy-wide average based on a dust-based technique by \citet{denBrok2023} is represented by the blue horizontal line. The red horizontal line represents our LTE-derived average $\alpha_{\rm CO}$. }
    \label{fig:aco_nonLTE}
\end{figure}


\begin{figure*}
    \centering
    \includegraphics[width=0.95\textwidth]{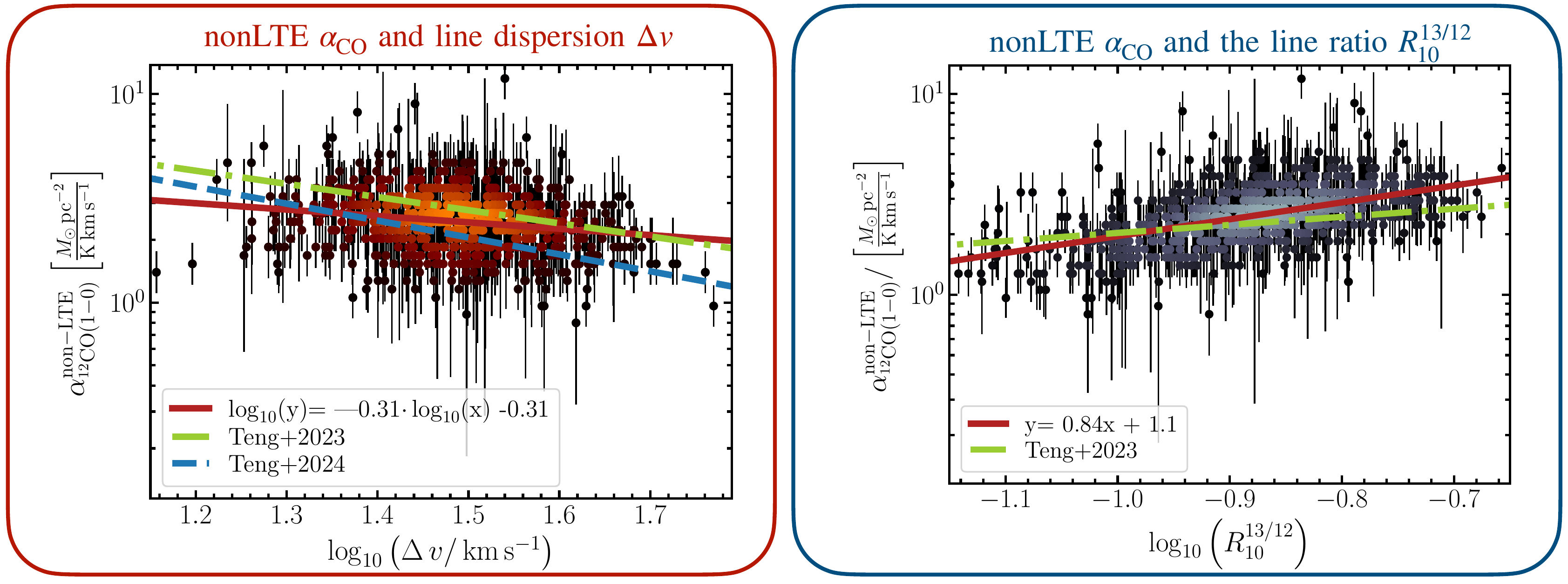}
    \caption{{\bf Analysis of the CO-to-H$_2$ conversion factor}  (\textit{Left}) The non-LTE-derived $\alpha_{\rm ^{12}CO(1-0)}$ correlated with the $^{12}$CO(1-0) line dispersion. The red line indicates a regression fit. For reference, we also indicate the \citet{Teng2023} and \citet{Teng2024} regression fits (after accounting for differences in the assumed CO-to-H$_2$ abundance values) with the blue and orange-dashed lines, which were obtained using a sample of 3 barred and 12 nearby star-forming galaxies, respectively. (\textit{Right}) The non-LTE-derived $\alpha_{\rm ^{12}CO(1-0)}$ correlated with the observed $^{13}$CO-to-$^{12}$CO ratio, $R_{10}^{13/12}$. The red line again shows a regression fit to the measurements of M51.  }
    \label{fig:aco_analysis}
\end{figure*}

From the non-LTE approach, we derive an estimate of the $^{12}$CO(1-0) optical depth. We expect that the $R_{10}^{13/12}$ line ratio traces a combination of changes in the optical depth and changes in the $^{12}$CO-to-$^{13}$CO abundance ratio.  To test the degree the ratio traces changes in optical depth, we plot the optical depth against the $R_{10}^{13/12}$ line ratio in \autoref{fig:optical_depth}.   Based on a linear regression fit (in log-log space) we find a mild but significant positive correlation ($R_p=0.41$, $p{\ll}0.05)$. The trend has a slope of $0.55\pm0.2$ for the correlation between $\log_{10}\left(R_{10}^{13/12}\right)$ and $\log_{10}\left(\tau_{^{12}\rm CO}^{\rm non-LTE}\right)$.  

From the non-LTE model fits, we obtain an estimate of the $^{12}$CO column density per line width, $N_{\rm ^{12}CO}/\Delta v$ (where $\Delta v$ corresponds to the line FWHM).  We furthermore can determine  a CO-to-H$_2$ conversion factor estimate for each grid point using the following formalism:

\begin{equation}\label{eq:aco_nonLTE}
\small
\begin{split}
    \alpha^{\rm non-LTE}_{\rm ^{12}CO(1-0)}\left[\frac{M_\odot\,\rm pc^{-1}}{\rm K\,km\,s^{-1} }\right]
    = \frac{1}{4.5\times 10^{19}}\cdot&\\
     \frac{1}{\left[\rm ^{13}CO/H_2\right]\cdot \left[\rm ^{12}CO/^{13}CO\right]}\cdot\frac{N(^{12}\rm CO)}{W^{\rm model}_{^{12}\rm CO(1-0)}}\cdot \phi,
\end{split}
\end{equation}
where the factor $4.5\times 10^{19}$ also considers helium. 
The accuracy of the derived CO-to-H$_2$ conversion factor also depends on the estimate of the $^{13}$CO-to-H$_2$ abundance ratio $\left[\rm ^{13}CO/H_2\right]$, which we convert into a $^{12}$CO abundance ratio by multiplying with the derived $\left[\rm ^{12}CO/^{13}CO\right]$ abundance ratio. We again rely on $^{13}$CO-to-H$_2$ abundance ratio of $1.7\times 10^{-6}$, which we employed already for the LTE analysis (see Section \ref{sec:ltemodeling}). Given our overall average $\left[\rm ^{12}CO/^{13}CO\right]$ abundance ratio of ${\sim}25$, this translates into a $^{12}$CO-to-H$_2$ abundance ratio of ${\sim} 0.4\times 10^{-4}$.
The final $\alpha_{\rm CO}$ estimate per line of sight is determined by taking the maximum of the normalized marginalized 1D PDF, similar to how we determine the non-LTE optical depth.

We plot the derived CO-to-H$_2$ conversion factor with respect to the galactocentric radius in \autoref{fig:aco_nonLTE}. The data points scatter around 0.3\,dex. Assessing the trend qualitatively, we observe that the non-LTE-based conversion factor does show a flat radial trend with an indication of a decrease towards the center at radii smaller than $r_{\rm gal}{<}0.5$kpc. 
Taking from \autoref{tab:nonLTE results}, we find the center ($r{<}20''$) to have an average of $\langle\alpha_{\rm ^{12}CO(1-0)}\rangle_{\rm center} = 2.7^{+0.8}_{-0.6}\,\frac{M_\odot\,\rm pc^{-2}}{\rm K\,km\,s^{-1}}$. Outside the center at larger radii, the value is slightly smaller with
 {$\langle\alpha_{\rm ^{12}CO(1-0)}\rangle_{\rm disk} = 2.3^{+0.9}_{-0.8}\,\frac{M_\odot\,\rm pc^{-2}}{\rm K\,km\,s^{-1}}$.} 

\begin{table}
        \centering
        \caption{{\bf Comparing LTE and non-LTE-based results} The mean and 16$^{\rm th}$-to-84$^{\rm th}$ percentile range of the overall distribution of values presented in \autoref{fig:comp_model} are provided.}
        \label{tab:comp_LTE_nonLTE}
        \begin{tabular}{l | c c}
        \hline 
         Parameter & LTE-based & non-LTE-based  \\ \hline \hline 
         $\langle\log_{10}\left(\tau_{^{13}\rm CO})\right\rangle$& ${-}0.9^{+0.6}_{-0.5}$ &${-}0.3^{+0.3}_{-0.3}$\\ 
         $\langle\log_{10}\left(N(^{13}\rm CO)/\rm cm^{-2}\right)\rangle$& $15.8^{+0.4}_{-0.3}$&$16.0^{+0.2}_{-0.2}$  \\\hline
    \end{tabular}

\end{table}

In comparison, on average, the non-LTE based values are systematically larger by  0.3\,dex (see \autoref{eq:aco_LTE}). We also note that the average value of $\langle\alpha_{\rm ^{12}CO(1-0)}\rangle = 3.1\,\frac{M_\odot\,\rm pc^{-2}}{\rm K\,km\,s^{-1}}$ reported by \citet{Sandstrom2013} for the nearby star-forming galaxy population falls within the scatter of our non-LTE derived values. Moreover, \citet{denBrok2023} find an identical value of $\langle\alpha_{\rm ^{12}CO(1-0)}\rangle_{\rm den\,Brok{+}2023} = 3.2\,\frac{M_\odot\,\rm pc^{-2}}{\rm K\,km\,s^{-1}}$ for the central radial 1\,kpc region, which is again the margin of scatter which we find at high angular resolution with the non-LTE CO isotopologue line modeling.

In \autoref{fig:aco_analysis}, we contrast the conversion factor to the molecular gas velocity dispersion as traced by the FWHM of the $^{12}$CO(1-0) line width. We find a weak ($R_p=-0.1$), but significant ($p{\ll}0.05$) negative trend between the two quantities. For reference, we plot the galaxy-wide relation found by \citet{Teng2024} for a sample of 12 nearby galaxies, who report a slope of $-0.81$ using kpc-scale dust $\alpha_{\rm ^{12}CO(1-0)}$ measurements and 150-pc-scale $\Delta v$. For this comparison, we adjust the literature values such that the CO-to-H$_2$ abundance ratio matches. Based on the average derived $^{12}$CO-to$^{13}$CO abundance ratio of 25 and the assumed $^{13}$CO abundance of $1.7\times 10^{-6}$, we find an average $^{12}$CO abundance of $4\times10^{-5}$. To derive $\alpha_{\rm ^{12}CO(1-0)}$ using CO isotopologue measurements, \citet{Teng2023} assume a $^{12}$CO abundance of $3\times10^{-4}$ for their sample of barred galaxy centers. In contrast, \citet{Teng2024} do not assume any abundance for their fit based on dust $\alpha_{\rm ^{12}CO(1-0)}$ measurements, but find that in order to match the \citet{Teng2023} results, the overall abundance appears to be $1.5\times 10^{-4}$. Therefore, we  {scaled} these trends to match our estimated CO-to-H$_2$ abundance. Using this  {scaling}, we find that the trends described by \citet{Teng2023} and \citet{Teng2024} match well the distribution of values we find in M51.

In the right panel of \autoref{fig:aco_analysis}, we show the trend with $R_{10}^{13/12}$. We find a mild ($R_p{=}0.3$) and significantly ($p{\ll}0.05$) positive correlation.  A positive correlation is expected under the explanation that the $R_{10}^{13/12}$ ratio reflects changes in the optical depth, which in turn drive variation in $\alpha_{\rm CO}$. \citet{Teng2023} also report a positive correlation with the $^{12}$CO-to$^{13}$CO line ratio. We note that in their study, they correlate the conversion factor with the ratio for the 2-1 instead of the 1-0 transitions (i.e., $R_{21}^{13/12}$ instead of $R_{10}^{13/12}$).  For reference, we plot the trend reported by \citet{Teng2023} in the panel as well (after adjusting for the difference in the CO-to-H$_2$ abundance, but not accounting for the fact that they provide it as a function of $R_{21}^{13/12}$ instead of $R_{10}^{13/12}$).  Despite the use of different CO isotopologue transitions, the trend matches well our derived distribution of $\alpha_{\rm CO}$. 

%
%
\section{Discussion } \label{sec:disc}
\subsection{Interpreting Global CO Isotopologue Line Ratio Trends}
The implications of the varying CO isotopologue ratios have been qualitatively assessed in M51 at kpc-scale in \citetalias{denBrok2022}. Generally, we differentiate between two types of line ratios: 

(a) line ratio of two different rotational-$J$ transitions for the same CO isotopologue species and 

(b)) line ratio of two different CO isotopologue species at the same rotational-$J$ transition.  

Case (a) line ratios probe excitation conditions since the individual line transitions each trace the column density (adjusted by the respective optical depth) for their respective excited level population \citep[e.g.,][]{Wilson1997,Narayanan2014, Penaloza2017}. In the case of type (b) ratios, the particular reason for variation within and across galaxies can be manifold. It also depends on whether the individual lines that form the ratios are optically thick or thin. The change in ratios can thus be linked to (i) differences in the excitation conditions for only one of the CO isotopologue species, (ii) changes in the abundance of the $^{13}$CO and C$^{18}$O species relative to $^{12}$CO, and (iii) changes in the optical depth. A more detailed description on the particular causes and drivers for CO isotopologue line ratio variation is given in \citet{Davis2014}. In the following sections, we aim to investigate the potential drivers for case (b) each separately and assess their impact based on our observed line ratio trends (\autoref{sec:ratiotrend}) and the insights gained from the CO isotopologue line modeling assuming LTE conditions (\autoref{sec:ltemodeling}) and with \texttt{RADEX} (\autoref{sec:radexmodeling}).
\subsubsection{Differences in the Excitation of the Lines}
At kpc-scale, \citetalias{denBrok2022} ruled out differences in the $^{12}$CO and $^{13}$CO excitation for M51 based on the fact that the $R_{10}^{13/12}$ and $R_{21}^{13/12}$ showed the same trend across the galaxy. 
Similarly, we observe that both $R_{10}^{13/12}$ and $R_{21}^{13/12}$ show a decreasing trend with galactocentric radius and an increasing trend with SFR surface density (see \autoref{fig:rgal_ratio}). Therefore, we conclude that differences in the excitation conditions between the different CO isotopologue species do not have a noticeable effect on the global line ratio trends we see in M51.

\subsubsection{Differences in the Relative Abundances}
We see evidence for changes in the relative abundances most clearly when looking at the line ratio of two optically thin lines, such as the transitions of $^{13}$CO and C$^{18}$O. Since both line transitions generally remain optically thin, we can link their brightness  {,to first order,} to the subsequent column density. The line ratio therefore gives us a direct measure of the relative abundances of the CO isotopologue species. In the ISM, we differentiate between three major potential drivers of CO isotopologue abundance variations. Here we present a short overview of the potential drivers of isotope abundance variation in question:
\begin{enumerate}
    \item {\bf Selective Photodissociation}\\
    The less abundant $^{13}$CO and C$^{18}$O species will be preferentially photodissociated in the presence of stellar populations with O and B stars \citep{vanDishoek1988}. The reason for this is a lower self-shielding (due to lower abundances). Also differences in the molecular structure cause different photodissociation rates, though this effect is mainly negligible for $^{13}$CO and only marginal for C$^{18}$O with rates that differ from $^{12}$CO by 0.1\% and 8\% respectively \citep{Visser2009}. Therefore, if selective photodissociation is the main driver for CO isotopologue ratio variation, we expect a strong link with star formation surface density, as OB stars are only short-lived.
    \item {\bf Chemical Fractionation}\\
    If the conditions are met, the $^{13}$CO abundance will increase due to the exchange reaction \citep{Keene1998}
    \begin{equation}
        {\rm ^{12}CO + ^{13}C^{+}} \rightarrow {\rm ^{13}CO + ^{12}C^{+}} .
    \end{equation}
    This reaction is exothermic. As the inverse reaction is endothermic it will be suppressed under cold ISM conditions (${<}$10\,K). As a result, the $^{13}$CO abundance will increase relative to $^{12}$CO.  
    Since $^{13}$CO emission is optically thin, the increase in its relative abundance to $^{12}$CO will affect the subsequent line ratios. We note, however, that in order for this effect to be important, most of the $^{13}$CO emission must be coming from gas which still contains significant $C^+$. So in the limit where CO dominates, fractionation will be limited and  changes in the $^{13}$CO/$^{12}$CO ratio abundance  vanishes. From theoretical considerations in turbulent clouds, \citet{Szucs2014} find that in turbulent clouds, the effect is most significant in lower column densities around $N(^{12}\rm CO){\sim}10^{16}\,cm^{-2}$, and becomes negligible at higher column densities beyond $N(^{12}\rm CO){>}10^{18}\,cm^{-2}$ (with the corresponding values for $^{13}$CO being $N(^{13}\rm CO){\sim}10^{14}-10^{16}\,cm^{-2}$). 
    \item {\bf Selective Nucleosynthesis}\\
    The relative CO isotopologue abundances are tracing the star formation history as different stellar populations lead to different C, N, and O isotope enrichment of the ISM. For instance, massive stars and their supernovae at the end of their stellar lives enrich the ISM mainly with $^{12}$C and $^{18}$O. The $ ^{13}$C isotope, in contrast, is a byproduct of the Bethe–Weizsäcker cycle \citep{Bethe1939}.   Intermediate-mass stars on the asymptotic giant branch inject the $ ^{13}$C isotope into the ISM via stellar winds \citep{Sage1991}. Therefore, in the case of nucleosynthesis as main driver, differences in the underlying stellar populations (massive or low-mass stars) will lead to different CO isotopologue line ratios, with larger $^{13}$CO abundances (relative to $^{12}$CO and C$^{18}$O) where intermediate-mass stars could inject $ ^{13}$C, while $^{12}$CO and C$^{18}$O abundances increase in the presence of younger stellar populations (i.e. at higher SFR surface densities).
\end{enumerate}

Qualitatively, we can assess the relevance of the aforementioned abundance variation as the main driver based on the sense of the observed line ratio trends with radius or star formation rate surface density. 
The $R_{10}^{13/12}$ line ratio shows an increasing trend with star formation surface density (see bottom center panel in \autoref{fig:rgal_ratio_2}). This suggests that selective photodissociation is unlikely the dominant driver of abundance variation since we would expect an opposite trend as more $^{13}$CO is photodissociated in the presence of O and B stars (as traced by higher SFR surface densities). In contrast, the trend agrees with both, selective nucleosynthesis and chemical fractionation as key drivers in case of abundance variation. Selective nucleosynthesis could lead, under the assumption of inside-out star formation history, to the observed trend as we expect regions with active star formation to have had more time to build stars and replenish the ISM with $^{13}$C ions from intermediate-mass stars. 
For chemical fractionation, using  our  median $N(^{13}\rm CO)$ from the LTE analysis ($\log_{10}N(^{13}\rm CO) = 15.6$), the models from \citet{Szucs2014} predict chemical fractionization only to be a 20-30\% effect.
Also the increasing $R_{10}^{13/12}$ trend with the SFR surface density (top center panel in \autoref{fig:rgal_ratio_2}) suggests that chemical fractionation is not the main driver, as we would expect the opposite trend as $^{13}$CO abundance increases at lower temperatures (i.e. lower SFR surface densities). We note that the slight upturn observed  at the lowest SFR surface densities in the plot is more likely related to signal-to-noise effects. 

The non-LTE modeling results allow us to be more quantitative about the observed abundance variations: We find overall very low $\left[^{12}\rm CO/^{13}CO\right]$ abundance ratio values of ${\sim}40$ across the galaxy and only marginally lower values of ${\sim}38$ toward the center or disk. These values match observational $\left[^{12}\rm CO/^{13}CO\right]$ abundances, which range from 30--60 (see Fig. 14 in \citealt{Szucs2014}, which compiles observational measurements from \citealt{Liszt1998}). Also,  already earlier studies of M51 have noted abundance ratios similar to ours. For instance, \citet{Schinnerer2010} assume an abundance ratio of 30 when studying the molecular gas in certain pointings along the southern spiral arm of M51. 
Furthermore, the low variation that we find in the abundance ratio is a strong indicator that abundance variation is not a main driver for changes in the $R_{10}^{13/12}$ line ratio.



\subsubsection{Changes in the Optical Depth}
 Our \texttt{RADEX} single-zone modeling shows a variation in the $^{12}$CO optical depth. This variation is correlated to the $R_{10}^{13/12}$ line ratio (see \autoref{fig:optical_depth}).
Indeed we expect changes in the optical depth, particularly of $^{12}$CO as it will vary proportionally with the gas surface density of the cloud and inverse proportionally with the kinetic temperature of the cloud and the gas velocity dispersion \citep{Paglione2001}.  Again, first assessing the line ratio trends qualitatively, we can see that the $R_{10}^{13/12}$ line ratio trend with SFR surface density (top middle panel in \autoref{fig:rgal_ratio_2}) agrees with expected changes in optical depth, as higher SFR surface densities correspond to higher kinetic temperatures and larger velocity dispersion (due to stellar heating of the gas). Such a relation between the SFR surface density and optical depth has been reported by previous studies as well \citep{Narayanan2014}. Moreover, the sharply decreased $R^{13/12}$ ratio toward the center (see \autoref{fig:rgal_ratio}) is consistent with lower $\tau_{^{12}\rm CO}$ in centers, and the negative correlation with FWHM is consistent with the scenario that dynamical effects decrease $\tau_{^{12}\rm CO}$ and increase the $^{12}$CO intensity and thus lowering $R^{13/12}$ \citep{Teng2024}.

\subsubsection{A combination of different drivers}
The derived CO isotopologue line ratio variation in combination with the non-LTE derived optical depth and isotopologue abundance ratio imply a combination of drivers for different parts of the galaxy. For instance, our results imply an impact of selective nucleosynthesis towards the center. We note that we would expect to see this also reflected in the non-LTE derived abundance ratio [$^{12}$CO/$^{13}$CO]. However, our detections are too limited in the central region to draw any meaningful conclusions.  

\subsection{The CO-to-H$_2$ conversion factor}

In \autoref{sec:ltemodeling} and \autoref{sec:radexmodeling}, we describe two approaches to estimating the CO-to-H$_2$ conversion factor. In the following, we will mainly focus on the non-LTE-derived $\alpha_{^{12}\rm CO(1-0)}$ values, and assess the implications of the observed trends. 

As is evident from \autoref{fig:aco_nonLTE}, we  detect a flat radial CO-to-H$_2$ conversion factor trend at $r_{\rm gal}{>}0.5$\,kpc and a decrease towards the center. However, due to the faintness of the CO isotopologues at the center, we do not have any non-LTE-derived values at $r{\footnotesize\lesssim}0.25$\,kpc that could solidify this apparent decrease towards the center, as the CO isotopologue emission becomes too faint.
Decreasing $\alpha_{^{12}\rm CO(1-0)}$ values towards the center are observed in other galaxies \citep{Teng2023, denBrok2023} and generally associated with lower $^{12}$CO emission opacities that increase the emissivity. Indeed, we see evidence of this in \autoref{fig:aco_analysis}, as there is a decreasing correlation of $\alpha_{\rm ^{12}CO(1-0)}$ and the line velocity dispersion. We note that also the LTE-derived conversion factor shows a decrease at $r{<}0.5$\,kpc (\autoref{fig:LTE_model2}). In contrast, however, the LTE-derived values show also a mild decrease towards larger radii, which is not observed for the non-LTE-derived radial trend. 
Overall, both approaches of determining the CO-to-H$_2$ conversion factor support the conclusion of a more uniform $\alpha_{\rm ^{12}CO}$ value across the disk. 

We note that, on average across our map, the LTE yields $\alpha_{\rm CO}$ values significantly lower ($\alpha_{\rm CO}^{\rm LTE}{\sim}1.0$) and non-LTE techniques values mildly below ($\alpha_{\rm CO}^{\rm nonLTE}{\sim}2.5$) those reported for similar star-forming galaxies based on dust observations. For instance, the disk-value reported for nearby galaxies by \citet{Sandstrom2013} is 3.1 $M_{\odot}\,\rm pc^{-2}/(K\,km\,s^{-1})$, which is 0.1\,dex higher than our galaxy-wide average for the non-LTE-derived value.  Our LTE-derived average lies even ${\sim}$0.5\,dex below the nearby galaxy average. Furthermore, \citet{denBrok2023} report a even higher dust-based $\alpha_{\rm ^{12}CO}(1-0)$  galaxy-wide average of $4.4 \pm 0.9\,M_{\odot}\,\rm pc^{-2}/(K\,km\,s^{-1})$. Finally, \citet{Schinnerer2010} also report a conversion factor similar to the Milky Way value based on an LVG modeling analysis using $^{12}$CO and $^{13}$CO at $4.5''$ in selected parts of the western arm of M51.  

But we also note that for some of the galaxies in the sample studied by \citet{Sandstrom2013}, comparable conversion factor values to our findings are reported as well (e.g., for NGC 3627, NGC 4569, NGC 4725, NGC 4736). 
And moreover, earlier studies found, in fact, CO-to-H$_2$ conversion factors below the Solar neighborhood value. For instance, \citet{Garcia1993} derived a value of $\alpha_{\rm ^{12}CO(1-0)}=1.6$ using an LVG modeling analysis based on $^{12}$CO and $^{13}$CO. When only using  $^{13}$CO, they report a conversion factor of $\alpha_{\rm ^{12}CO(1-0)}=1$, which is consistent to our LTE-derived value.  

\begin{figure}
    \centering
    \includegraphics[width=0.6\columnwidth]{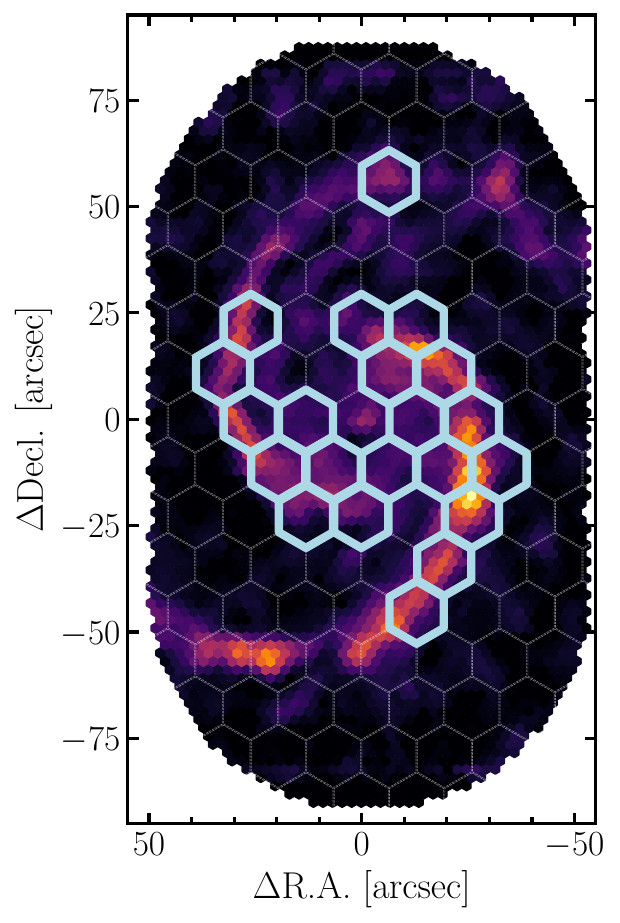}
    \caption{{\bf Deriving an estimate for the dust-to-gas ratio (DGR).}  The background illustrates the $^{12}$CO(1-0) moment-0 map at $4''$. The larger hexagonal grid indicates the map from \citet{denBrok2023} for which we have H\textsc{i} and dust mass measurements (at $26''$ resolution). For the blue marked hexagons, we have non-LTE $\alpha_{\rm CO}$ values for ${>}70\%$ in terms of the total $^{12}$CO(1-0) intensity from within that pixel. }
    \label{fig:dgr_analysis}
\end{figure}

To test whether our non-LTE conversion factor values are consistent with dust-based $\alpha_{\rm CO}$ measurements, we calculate the implied dust-to-gas ratio (DGR) based on the CO-to-H$_2$ conversion factor, the atomic mass surface density (from 21cm H\textsc{i} observations), and the dust mass surface density (from infrared SED-based measurements). Such data exists at coarser, 26$''$ resolution and have been presented in \citet{denBrok2023}.
We use the following prescription to obtain an estimate of the DGR on kpc-scale:
\begin{equation}
    {\rm DGR}= \frac{\Sigma_{\rm H\textsc{i}}+\langle\alpha_{\rm CO}\rangle^{26\arcsec}\cdot W_{^{12\rm CO(1-0)}}^{26\arcsec}}{\Sigma_{\rm dust}}
\end{equation}

In \autoref{fig:dgr_analysis}, we present the 4$''$ $^{12}$CO(1-0) observation of M51. We overplot the coarser $26\arcsec$ grid on which we have the H\textsc{i} and dust mass surface density observations.  
Therefore, we have to match our $\alpha_{\rm CO}$ derived at high angular resolution first to the coarser 26$\arcsec$ observations. To do this, we derive a $^{12}$CO(1-0) intensity weighted average of the $\alpha_{\rm CO}$ conversion factor per coarse grid point (i.e., $\langle\alpha_{\rm CO}\rangle^{26\arcsec}$ is the luminosity weighted $\alpha_{\rm CO}$). We can subdivide each 26$\arcsec$ into a set of $4\arcsec$ gridpoints for which we have an estimates of  $\alpha_{\rm CO}$.  We only compute a weighted average conversion factor (and consequently DGR) for 26$\arcsec$ gridpoints for which include in their 4$\arcsec$ sub-beams enough $\alpha_{\rm ^{12}CO(1-0)}$ values. As threshold, we set that the points for which we have a significant $\alpha_{\rm ^{12}CO(1-0)}$ measurement need to contribute $>70\%$ of the total $^{12}$CO(1-0) intensity from within this hexagonal gridpoint. 
Using this approach, we derive an average DGR  {for the inner $r{\lesssim}3\,\rm kpc$ region} of 0.012$\pm0.003$, which is within the range of the fiducial value of 0.01 commonly used in the literature.


In general, it is not uncommon to find disagreement between CO-to-H$_2$ values derived using different techniques. For instance,  dust-based CO-to-H$_2$ estimates do not rely necessarily on a $^{12}$CO abundance ratio, but on a dust-to-gas ratio, which also remains uncertain and can explain the 0.2\,dex offset to our non-LTE values. In contrast, we also note that the low values we measure within the disk at $r{\lesssim}3$\,kpc are actually in agreement with values reported for the centers ($r{\le}0.5$\,kpc) of nearby galaxies \citep{Israel2020, Teng2023}.


One major limitation of our non-LTE approach is that we only can use effectively four of the six CO isotopologue lines. The C$^{18}$O(2-1) emission is too weak and our SMA observations not sensitive enough to detect it across a more extended region in M51. And only including C$^{18}$O(1-0) will just allow us to constrain the C$^{18}$O abundance. However, doing the \texttt{RADEX} modeling fit with only four lines, we need to make assumptions on certain parameters, that ideally, we would want to leave free, such as the volume density width\footnote{The volume density width describes the range in volume densities within our beam.} or the beam filling factors. Therefore, to improve the constraints and accuracy of the non-LTE-derived conversion factor, either more sensitive C$^{18}$O(2-1) observations would be necessary, or higher-$J$ transitions of the $^{12}$CO and $^{13}$CO could be targeted to fit more free parameters.

\begin{figure}
    \centering
    \includegraphics[width=\columnwidth]{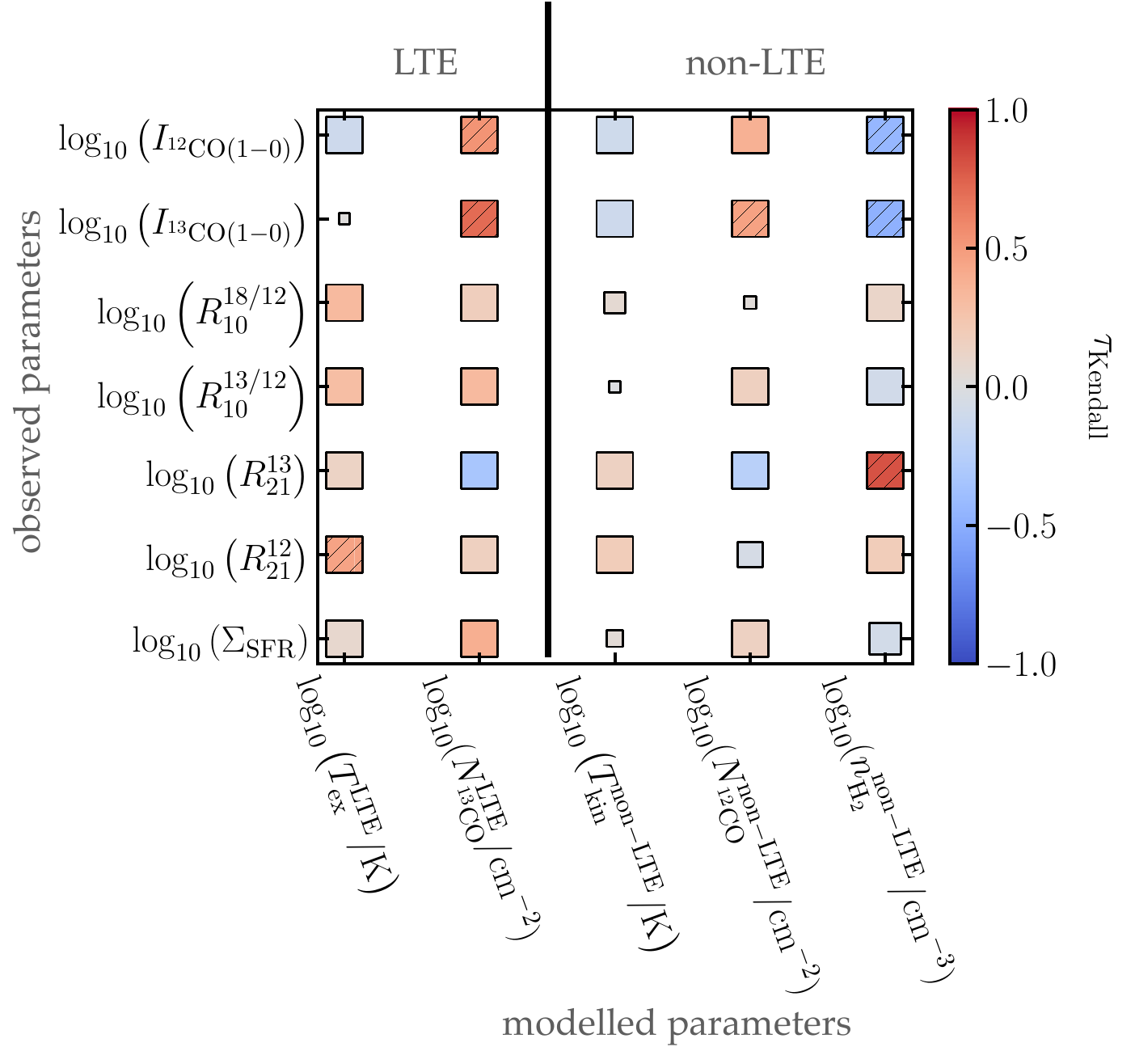}
    \caption{{\bf Correlation of LTE and non-LTE parameters.}  Each box represents the strength of the LTE or non-LTE derived parameters (on $x$-axis) with some key observables or galactic parameters (on $y$-axis). We note that we have scaled the non-LTE $^{12}$CO column density by the beam filling factor. Red indicates a positive and blue a negative correlation.  The correlation is quantified by the Kendall's $\tau$ correlation coefficient.  The size of the points indicates the inverse $p$-value (larger points show a significant correlation with $p{<}0.05$). Strong correlations, where $|\tau|>0.5$ are indicated by hatched markers. }
    \label{fig:corr_plot_all}
\end{figure}

 {
\subsection{Limitations of modeling assumptions}
We emphasize that both the LTE and non-LTE line fitting approaches depend on assumptions that can potentially induce bias and lead to inaccurate fitting results. The key assumptions we build into our models are (i) that the conditions can be described by a one-zone model, i.e. homogeneous conditions across a beam, (ii) all the CO isotopologue emission is originating from the same area/volume, and (iii) the overall CO abundance (in our case [$^{13}$CO/H$_2$]) stays constant.}

 {On limitation (i), assuming a one-zone model may oversimplify the complex structure of GMCs, particularly in environments where gradients in density, temperature, or chemical abundances exist on scales smaller than our beam size. Such complexities could result in underestimating or overestimating the derived physical parameters and explain part of the pixel-to-pixel scatter in our results. We emphasize, however, that such a description of the molecular gas using a single component only is widely used in the literature \citep[e.g., ][]{Ripple2013,Topal2014, Sliwa2017, Cormier2018, Teng2022, Wange2023}.} {In particular, when based on low-J $^{12}$CO and $^{13}$CO line emission, such a modeling approach remains viable for deriving beam-averaged conditions, as \citet{Leroy2017dens} find that only for tracers at higher critical density, the beam-averaged emission does indeed become sensitive to the particular sub-beam density distribution. }

 {The assumption that all CO isotopologues originate from the same region may also be problematic if selective photodissociation or chemical differentiation causes spatial segregation of different isotopologues. Galactic and extragalactic CO mapping studies indeed appear to show a spatial difference between the origin of $^{12}$CO and $^{13}$CO emission \citep[e.g.,][]{Goldsmith2008,Pety2017, denBrok2022}. However, stacking analyses usually reveal the presence of faint $^{13}$CO in all regions bright in $^{12}$CO, suggesting that the detection of this isotopologue is generally limited by sensitivity. This supports the use of an identical beam filling factor for the $^{12}$CO and $^{13}$CO isotopologues and the use of their ratios to trace variation in the overall molecular gas physical conditions. Furthermore, studies in the Orion B cloud complex also reveal a similar behavior of the $^{12}$CO, $^{13}$CO, and C$^{18}$O in terms of their variation and trends \citep[e.g.,][]{Pety2017, Tafalla2023} with the overall H$_2$ column density. This suggests that using a single-component approach to derive beam-averaged parameters remains reasonable. Only in the low-density envelope of molecular clouds the behavior of the different CO isotopologue lines becomes more uncertain. However, here, the intensities of the emission also drop, therefore limiting the impact on the beam-averaged line ratios. }

 {Finally, we again emphasize that in particular the derived $\alpha_{\rm CO}$ value depends critically inversely on the assumed $^{13}$CO-to-H$_2$ abundance ratio. We already discussed the selection and effect of this abundance ratio in Subsection \ref{sec:alpha_CO_nonLTE} and explained why the assumption of a constant abundance ratio is reasonable, at least for a limited region within M51. However, in essence, this means that our derived conversion factor values should be interpreted as $\alpha_{\rm CO}$ per abundance ratio. When comparing the conversion factor with results from other studies, the difference in assumed abundance ratio needs to be accounted for.}

 {
Therefore, while we emphasize that our modeling approach is subject to inherent limitations due to the set of assumptions (as are other studies in this field), we argue that they do not significantly compromise the robustness of our results. Nevertheless, these assumptions should be kept in mind when interpreting findings, and future work that resolves finer-scale structures or explores variable abundance ratios will be crucial in refining these models.}

\subsection{LTE vs. non-LTE line fitting}
In this study, we examined the emission of CO isotopologue lines using two approaches: LTE and non-LTE. However, both methods depend on certain underlying assumptions. In the ISM, assuming thermalization (i.e., ``LTE") is usually not accurate. Instead, to obtain a more precise description of the partition function, it is necessary to solve the equations of statistical equilibrium (i.e., non-LTE calculation). However, this requires detecting several fainter emission lines to determine the different degrees of freedom. In our study, we used four CO isotopologue lines to determine four chemical and physical parameters of the ISM, including temperature, density, column density, and relative isotopologue abundances.

In \autoref{fig:corr_plot_all}, we provide a direct comparison of the derived quantities from both the LTE and non-LTE approach with other key galactic observables. Strong correlations, which have a Kendall's $\tau$ correlation coefficient of $\tau_{\rm Kendall}{>}0.5$ are indicated by hatched marker symbols. 

Overall, we only find a few strong correlations ($\tau_{\rm Kendall}{>}0.5$) for the LTE-derived parameters. The $^{13}$CO column density scales significantly with both the $^{13}$CO(1-0) and $^{12}$CO(1-0) intensities. The correlation with $^{13}$CO(1-0) is expected as it follows from \autoref{eq:cdens}. The correlation with $^{12}$CO(1-0) follows as we do not find any strong variation of the $R_{10}^{13/12}$ (apart from a depression towards the very center). Finally, we also find a strong positive correlation of the $R_{21}^{12}$ with the excitation temperature, which is expected as the excitation temperature is directly 
proportional to the peak temperature ratio of $^{12}$CO. Previous studies have concluded that with LVG models, it is difficult to constrain any physical parameter within 0.5\,dex, with likely even larger uncertainties when accounting for flux calibration uncertainties of 10\% \citep{Tunnard2016}.
This could explain the large scatter for our LTE-derived parameters and why there is no significant correlations with key galactic observables.

The non-LTE derived parameters do also show only few strong correlations. We find a strong correlation between the $^{13}$CO(1-0) intensity and the $^{12}$CO column density, indicating that $^{13}$CO(1-0) is (under some caveats) a good tracer of the amount of molecular gas.  {As $^{13}$CO does generally not become optically thick, in contrast to $^{12}$CO (see \autoref{fig:comp_model}), it traces more directly the column density of gas, while the $^{12}$CO intensity is more susceptible to changes in opacity.} We also find significant negative correlations between the $^{12}$CO(1-0)  and $^{13}$CO(1-0)  intensities and the H$_2$ volume density. The negative correlation could indicate opacity effects, as at large densities, CO becomes more opaque  {(which follows from the definition of the optical depth, which is the integral along the line of sight of the absorption coefficient, which in turn scales with the volume density)}. The strong positive correlation of the volume density with the $R_{21}^{13}$ ratio, in contrast, reflects the impact of increasing density on the underlying excitation conditions.

In summary, with the set of four CO isotopologue emission lines at our disposal, the non-LTE-derived parameters do provide improved constraints only in the sense of a decreased scatter. In particular, the $^{13}$CO column density and opacity, as well as the CO-to-H$_2$ conversion factor, agree within the range of scatter with both approaches, at least  within the disk. In the center, where conditions become much different from the rest of the galaxy in terms of temperature, density, and dynamics, more emission lines are needed to accurately constrain key parameters such as temperature and density variation at sub-beam scales.
%
%
\section{Conclusion} \label{sec:conc}

We present a multi-CO isotopologue line analysis of M51 combining observations from multiple mm-interferometer large programs. This includes $^{12}$CO(1-0) data from PAWS, $^{13}$CO(1-0) and C$^{18}$O(1-0) from SWAN, and $^{12}$CO(2-1), $^{13}$CO(2-1) and C$^{18}$O(2-1) from the SMA M51 LP survey. All data have angular resolutions of ${\le}4''$, which corresponds to a spatial scale of ${\le}170\,$pc. In combination, we assess the CO isotopologue line ratios across  {the central $r{<}3$~kpc region of} M51 and perform LTE and non-LTE line modeling to assess variation in the molecular gas conditions across the different environments of M51.  In particular, we find the following.
\begin{enumerate}
    \item The $R_{21}^{12}$, $R_{21}^{13}$, $R_{10}^{13/12}$, $R_{21}^{13/12}$, and $R_{10}^{18/13}$ line ratios all show a mild decreasing radial trend for $r{>}0.5$\,kpc. In contrast, at $r{<}0.5$\,kpc, the trends differ. While $R_{21}^{12}$, $R_{21}^{13}$ significantly increase towards the center,   $R_{10}^{13/12}$, $R_{21}^{13/12}$, and $R_{10}^{18/13}$ show all a decrease.
    \item We also find an increasing trend of the CO isotopologue line ratios with SFR surface density. In contrast, no line ratio correlates with the $^{12}$CO line width, which traces the molecular gas velocity dispersion.

    \item Under LTE-assumptions, we use the $^{12}$CO emission to estimate the excitation temperature and $^{13}$CO(1-0) to estimate the $^{13}$CO optical depth and column density. Using a fiducial $^{13}$CO-to-H$_2$ abundance ratio of $1.7\times10^{-6}$, we derive a galaxy wide CO-to-H$_2$ conversion factor value at $r{\lesssim}3\,$kpc of ${\sim}0.8\,M_\odot\rm\,pc^{-2}/(K\,km\,s^{-1})$. This is a factor 5 below the fiducial value assumed for the MW local neighborhood of 4.3.

    \item We use a \texttt{RADEX} grid to model $T_{\rm kin}$, $n_{\rm H_2}$, $N_{\rm ^{12}CO}/\Delta v$, and $\left[^{12}\rm CO/^{13}CO\right]$. Using a $\chi^2$ minimization approach, we fit these parameters to the observed line intensities to find the best-fit molecular gas conditions.  We convert the derived $^{12}$CO column densities into an estimate of the CO-to-H$_2$ conversion factor. We find a nearly flat trend of $\alpha_{\rm ^{12}CO(1-0)}^{\rm non-LTE}\approx2.4\,M_\odot\rm\,pc^{-2}/(K\,km\,s^{-1})$, which 0.3\,dex larger than the LTE value.
\end{enumerate}

Contrasting LTE and non-LTE derived parameters, we note that the non-LTE approach produces lower scatter. The LTE approach also seems to underestimate the column density by a factor 2--3. For a more robust non-LTE approach, more high-$J$ CO isotopologue lines are required to obtain constraints also on the sub-beam density distribution and/or beam filling factor.
{Given the simplistic one-zone model, we emphasize that the LTE and non-LTE-derived values must be interpreted carefully when contrasting with Galactic or other extragalactic work using different modeling prescriptions.}

Overall, these results highlight the diverse nature of the molecular ISM in nearby galaxies with clear differences between the various galactic environments. The results highlight that when assessing the molecular gas properties using CO observations, we need to account for the particular galactic environment.
%
%

\section*{Acknowledgments}
This work was carried out as part of the PHANGS collaboration. {We thank the anonymous reviewer for their careful reading of the manuscript and their insightful feedback and suggestions.}
This work is based on observations carried out  under project ID M19AA with the IRAM NOEMA Interferometer. IRAM is supported by INSU/CNRS (France), MPG (Germany) and IGN (Spain).
We also use data from the Submillimeter Array which  is a joint project between the Smithsonian Astrophysical Observatory and the Academia Sinica Institute of Astronomy and Astrophysics and is funded by the Smithsonian Institution and the Academia Sinica. {We recognize that Maunakea is a culturally important site for the indigenous Hawaiian people; we are privileged to study the cosmos from its summit.}
J.d.B. and E.K. acknowledge support from the Smithsonian Institution as Submillimeter Array (SMA) Fellows. 
MJJD, AU and MQ acknowledge support from the Spanish grant PID2022-138560NB-I00, funded by MCIN/AEI/10.13039/501100011033/FEDER, EU.
ES acknowledges funding from the European Research Council (ERC) under the European Union’s Horizon 2020 research and innovation programme (grant agreement No. 694343).
K.S. and Y.-H. T. acknowledge funding support from the National Science Foundation (NSF) under grant No. 2108081. E.R. and Engineering Research Council of Canada (NSERC), funding reference number RGPIN-2022-03499.
PKH gratefully acknowledges the Fundação de Amparo à Pesquisa do Estado de São Paulo (FAPESP) for the support grant 2023/14272-4. 
SKS acknowledges financial support from the German Research Foundation (DFG) via Sino-German research grant SCHI 536/11-1.
M.C. gratefully acknowledges funding from the Deutsche Forschungsgemeinschaft (DFG, German Research Foundation) through an Emmy Noether Research Group (grant number CH2137/1-1). COOL Research DAO is a Decentralised Autonomous Organisation supporting research in astrophysics aimed at uncovering our cosmic origins. 
H.A.P. acknowledges support from the National Science and Technology Council of Taiwan under grant 110-2112-M-032-020-MY3. 

\vspace{5mm}
\facilities{Submillimeter Array (SMA), Institut de radioastronomie millimétrique (IRAM) telescopes, Extended Very Large Array (EVLA)}

\software{astropy \citep{2013A&A...558A..33A,2018AJ....156..123A},  
          SpectralCube \citep{Ginsburg2015}} , PyStructure \citep{PyStructure_v3}

\appendix
\section{SMA Data Reduction}
\label{app:dataReduxSMA}
We use observations from the Submillimeter Array M51 Large Program  (2016B-S035; PI: K. Sliwa). In the following, we briefly describe the observations, data calibration, imaging, and short-spacing correction that we used to produce the data cubes. 
\subsection{SMA Observations and Data Calibration}
The SMA observed M51 using three different configurations split into 14 different tracks over the course of 2017. We note that if the sensitivity requirements of the individual tracks were not achieved, we obtained additional, further tracks. \autoref{tab:obs_summary} provides an overview of all the observations of this program. The main tracks 1--6 are done with the SMA in compact (COM) configuration. These consist of 55 pointings such that the resulting field of view matches the one from PAWS (see \autoref{fig:outline}). Track 8 consists of SMA extended (EXT) configuration observations. The remaining tracks 9--14 are done using the SMA subcompact (SUB) configuration. For the SUB configuration observations, the observations consist of more pointings, resulting in a larger field of view (see the SMA outline shown in \autoref{fig:outline}). Apart from track 7, all tracks could be observed under sufficient weather conditions with nightly-average $\langle\rm pwv\rangle{\lesssim}4$\,mm.

The SMA data calibration was performed using the MIR software package\footnote{\url{https://lweb.cfa.harvard.edu/~cqi/mircook.html}}. Flux calibrators were, depending on availability during the night, either Titan, Jupiter, or Saturn. For the bandpass, the SMA observed either 3c84 or 3c454.3. As gain calibrator two sources were observed throughout the night (two of the following: 1419+543, 1310+323,  and 1153+495). 

\begin{deluxetable}{rlcccc}
    \tablecaption{Summary of observations.  \label{tab:obs_summary}}

    \tablehead{\multicolumn{2}{c}{Track} &\colhead{Date}&\colhead{configuration}&\colhead{$\langle\rm pwv\rangle$}&\colhead{on-source time}\\  &  & \colhead{[mm/dd/yy]} & & \colhead{[mm]} & \colhead{[h]}} 

\startdata 
         1&a&04/17/2017&COM&0.8&3.4 \\
         \multirow{2}{*}{2} &a&04/19/2017&COM&2.1&5.6 \\
         &b&05/04/2017&COM&1.5&5.2\\
         3&a&04/29/2017&COM&1.8&5.3 \\
         4&a&04/15/2017&COM&1.1 &5.5\\
         \multirow{2}{*}{5} &a&04/14/2017&COM&4.1&3.3 \\
         &b&04/26/2017&COM&4.3&4.9\\
         6&a&04/13/2017&COM&2.6 &3.2\\
         7&a&05/03/2017&COM&7.1&-- \\
         \multirow{3}{*}{8} &a&02/22/2017&EXT&2.0 & 8.4 \\
         &b&02/27/2017&EXT&3.3&2.9\\
         &c&03/07/2017&EXT&3.1&4.6\\
         9&a&01/24/2017&SUB&1.6 &3.2\\
         10&a&02/09/2017&SUB&1.3 &3.4\\
         11&a&05/28/2017&SUB&2.1 &3.0\\
         \multirow{2}{*}{12} &a&02/13/2017&SUB&2.6 &2.4\\
         &b&02/15/2017&SUB&1.6&3.4\\
          \multirow{3}{*}{13} &a&05/22/2017&SUB&3.3&3.2 \\
         &b&05/27/2017&SUB&1.1&3.0\\
         &c&06/01/2017&SUB&2.8&3.1\\
         14&a&02/10/2017&SUB&2.3 &3.5\\
\enddata
\end{deluxetable}

\subsection{Imaging}
We first convert calibrated visibilities into a measurement set readable in \texttt{CASA} \citep{McMullin2007}. This step is needed so that we can use the \textit{PHANGS-ALMA imaging pipeline} \citep{Leroy2021}\footnote{\url{https://github.com/akleroy/phangs_imaging_scripts/}}.
We add scan intents to the SMA data to mimic the ALMA format.
The pipeline first transforms the visibility data to a user-given channel width and frequency range, that corresponds to the line plus line-free channels. 
Unlike for ALMA data, we avoid reweighting the data as the SMA's data weights are calibrated to system temperature, accurately reflecting the measurement uncertainty.
Next, the continuum emission is subtracted. Then a two-stage cleaning process is started using the task \texttt{CASA tclean}. The task \texttt{tclean} was run iteratively to force numerous large cycles (improving the accuracy of the deconvolution). The first stage consists of  a multi-scale cleaning down to ${\sim}4\sigma$. For the second stage, a single-scale cleaning is performed down to ${\sim}1\sigma$. 
In addition, as part of the second stage, a clean mask is computed at each iteration. This clean mask is constructed using a watershed algorithm that extracts all ${>}2{\sigma}$ pixels connected to ${>}4\sigma$ peaks.

Since the SUB and COM observations have a different number of pointings, and CASA cannot deal with a significant spatial variations in the PSF within a single mosaic, we removed pointings in the SUB field that did not have a matching pointing in the COM observations. Furthermore, the C$^{18}$O(2-1) line falls within the edge of chunks (spectral windows) 1 and 2 of the lower sideband, so we stitched together the two sidebands using the \texttt{CASA concat} and \texttt{mstransform} functions.

\subsection{Short-Spacing Correction and Post-Processing}
\begin{figure}
    \centering
    \includegraphics[width=0.7\textwidth]{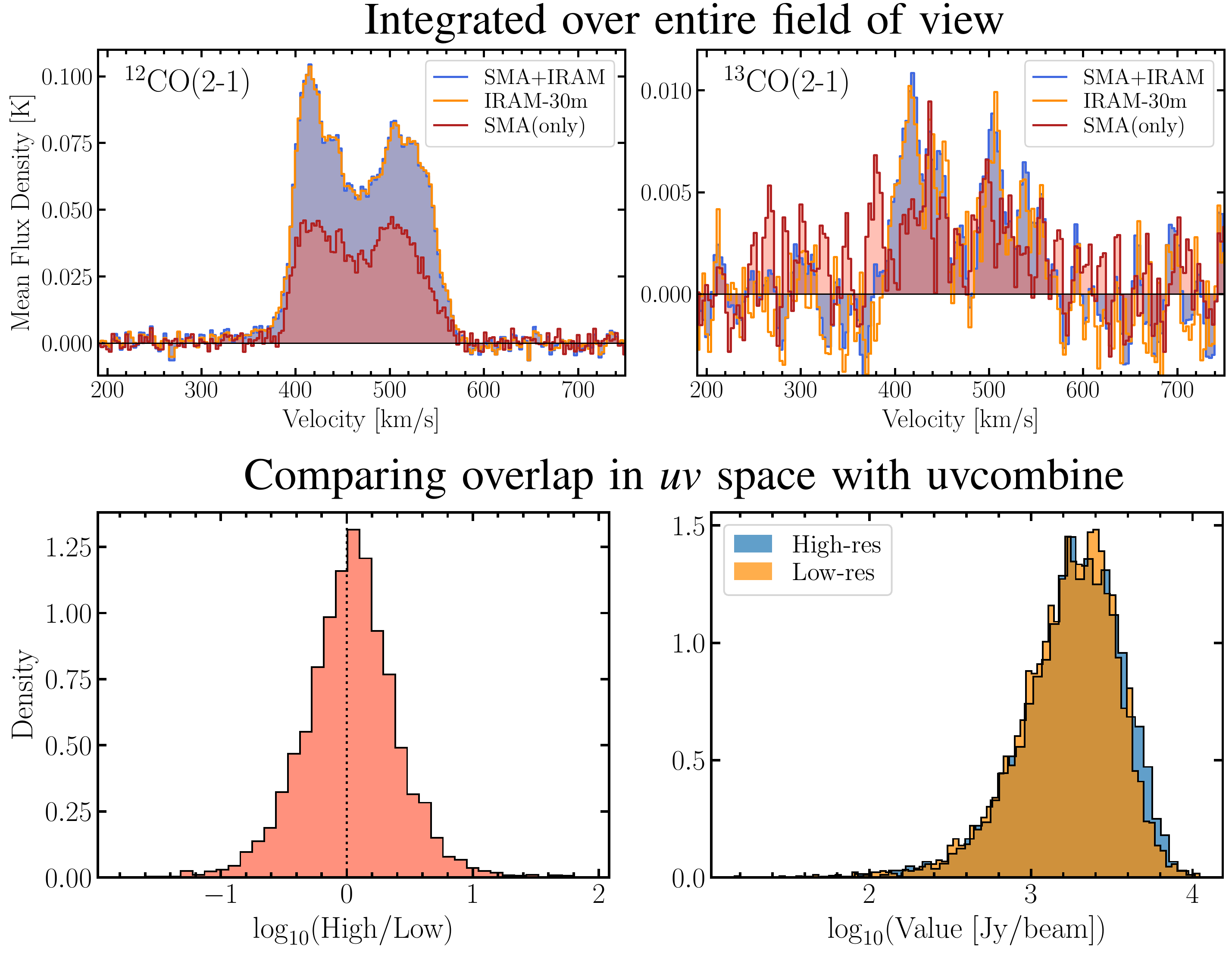}
    \caption{{\bf Short-spacing correction analysis.} The top panels show the $^{12}$CO(2-1) (\textit{top left}) and $^{13}$CO(2-1) (\textit{top right}) spectra, averaged over the entire field of view. We obtain the red spectrum when using only the SMA SUB+COM cube. After short-spacing correction (SMA+IRAM), we obtain the blue spectrum. For $^{12}$CO(2-1), it is evident that half of the emission is missing in the interferometric data. The lower panels show the amplitudes in the overlapping $uv$ space of the IRAM and SMA data (ranging from $13''-28''$) ratio of the amplitudes (bottom left), and the amplitude distribution (bottom right) for the high (SMA) and low (IRAM 30m) resolution data separately.}
    \label{fig:ssc_analysis}
\end{figure}

We use the \textit{PHANGS-ALMA post-processing pipeline}, which is an extension of the \textit{PHANGS-ALMA imaging pipeline} for the short-spacing correction. The post-processing routine first uses the \texttt{CASA feather} routine for the short-spacing correction. We use the IRAM 30m data from CLAWS \citepalias{denBrok2022} for the single-dish data. Then the pipeline corrects the resulting images for the primary beam attenuation and converts them to Kelvin units.

In \autoref{fig:ssc_analysis}, we present a rudimentary analysis of the short-spacing corrected data. The top two panels show the averaged spectra over the entire field of view. For $^{12}$CO(2-1), we find that 50\% of the flux appears to be missing in the interferometric data. This degree of diffuse emission is similar to what \citet{Pety2013} report for the $^{12}$CO(1-0). To verify that this discrepancy is not the result of flux calibration uncertainties, we match the amplitudes of the visibilities in the overlapping $uv$ space of the SMA and IRAM 30m data using the \texttt{uvcombine} package \citep{Koch2022}. The $uv$ overlap extends from ${\sim}12''$ (resolution of IRAM 30m data) to ${\sim}28''$ (largest recoverable scales of the SMA in SUB configuration). As is evident from the distribution of the visibilities in the overlapping $uv$ space, the visibilities agree within the noise variation (ratio of $1.09\pm0.07$ for interferometric versus single dish flux), indicating consistent flux calibrations for the two datasets.

\begin{deluxetable}{c c c c c c c c c c}
    \tablecaption{Table of \texttt{RADEX} single-zone modeled line intensities. The full table is avaliable in machine-readable form}
    \label{tab:radex_grid}

    \tablehead{\colhead{$T_{\rm kin}$}&\colhead{$n_{\rm 0, H_2}$}&\colhead{$N_{\rm CO}/\Delta v$}&\colhead{$\left[^{12}\rm CO/^{13}CO\right]$}  &\colhead{$\phi$}&  \colhead{$\tau_{^{12}\rm CO}$} & \colhead{$\tau_{^{13}\rm CO}$} &\colhead{$W({\rm ^{12}CO(1-0)}$)}&\colhead{...}\\ \colhead{[K]}&\colhead{[cm$^{-3}$]}&\colhead{[cm$^{-2}$/km/s$^{-1}$]}& \colhead {} & \colhead {} & \colhead {} &    \colhead {}  & \colhead{[K\,km\,s$^{-1}$]} &} 
\startdata 
         4.0&100.0&$5\times10^{14}$&20 & 0.01 & 0.571 	&0.0274 &	0.0035 \\
         4.0&100.0&$7.925\times10^{14}$&20&0.01 & 0.9014 	& 0.0435 & 0.0057\\
         4.0&100.0&$1.256\times10^{15}$&20& 0.01 &1.419 &	0.0689 &0.0090 	\\
         4.0&100.0&$1.991\times10^{15}$&20&0.01&2.2240 & 0.1091  & 0.0142\\
         4.0&100.0&$3.155\times10^{15}$&20&0.01&3.463 & 0.1728 	&0.0218 	\\
         $\vdots$&$\vdots$&$\vdots$&$\vdots$&$\vdots$&$\vdots$&$\vdots$&$\vdots$&$\vdots$ \\
         100 & $10^5$&$1.991\times10^{18}$ & 80& 1 &  5.640 	&0.0649& 	1528\\
         100 & $10^5$&$3.155\times10^{18}$ & 80&1 &  8.904	&0.1021 &	1534 \\
         100 & $10^5$&$5.0\times10^{18}$ & 80&1 &  14.07 	&0.1612 &	1535 \\
\enddata
\end{deluxetable}





\section{Extent of significant line ratio detections}
\label{app:detection_extension}

\begin{figure*}
    \centering
    \includegraphics[width=0.95\textwidth]{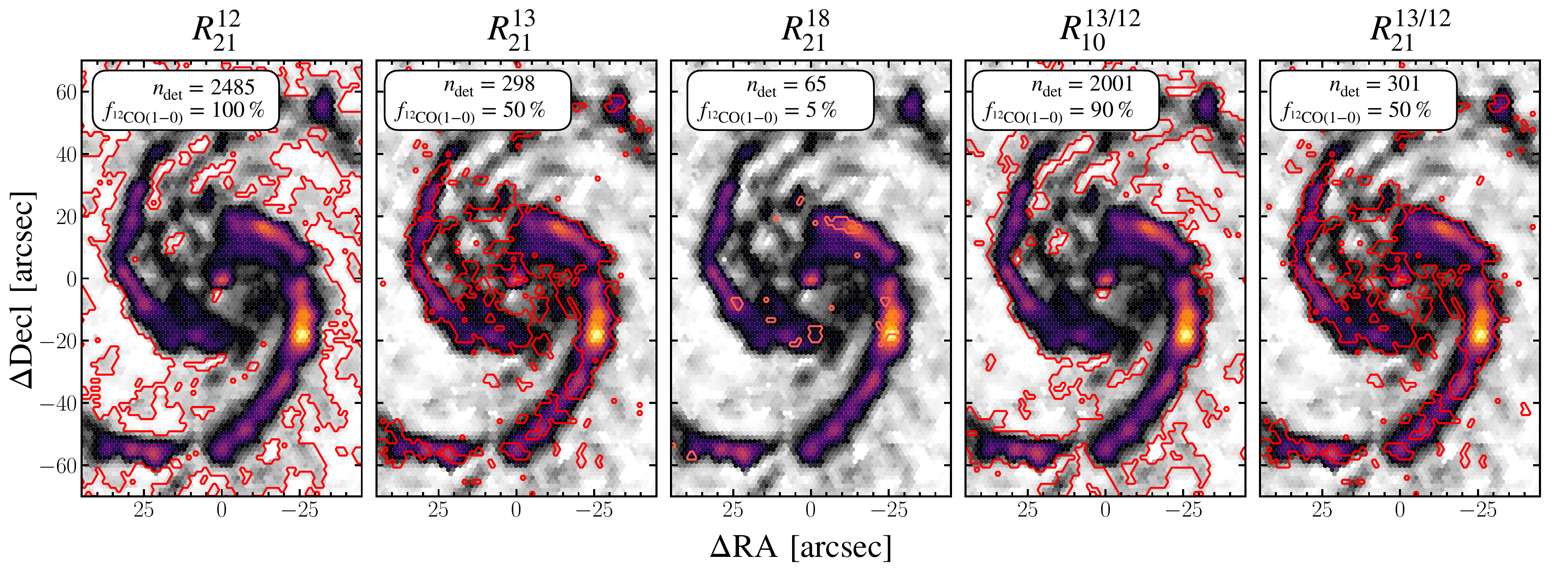}
    \caption{{\bf Significant (5$\sigma$) line ratio detection}.  {Each panels illustrates the sightlines per line ratio for which both lines are significantly detected at $\rm S/N{>}5$ by the red contour. For reference, each panel also lists the number of sightlines and the percentage of $^{12}$CO(1-0) flux contained for the selected sightlines in the top left corner. The sightlines are colored by the CO(1-0) moment-0 to show M51's morphology.}}
    \label{fig:ratio_detect}
\end{figure*}

 {For the line ratio analysis, we focus primarily on sightlines for which both lines forming the ratio are significantly (i.e. $\rm S/N{>}5$)  detected. Since the sensitivity of the different observations and the strength of the different lines varies, the number of significant detections changes. In \autoref{fig:ratio_detect}, we illustrate the spatial distribution for the significantly detected sightlines per line ratio for the ratios presented in \autoref{fig:rgal_ratio} and \autoref{fig:rgal_ratio_2}. }
\section{\texttt{RADEX} non-LTE Grid}
\label{app:radex_grid}
For our non-LTE line modeling approach, we construct a parameter grid using \texttt{RADEX} \citep{vanTak2007}. This grid contains the simulated integrated intensities of the CO isotopologues for the various permutations of free parameters. The free parameters are the kinetic temperature, $T_{\rm kin}$, the H$_2$ volume density, $n_{\rm H_{2}}$, the CO column density per line width, $N_{\rm CO}/\Delta v$, and the CO isotopologue abundance ratio $\left[^{12}\rm CO/^{13}CO\right]$. The range of parameter values used to generate the grid are described in \autoref{tab:mode_param}. We fix the $^{12}$CO line width, $\Delta v$, to 15\,km\,s$^{-1}$. We provide the entire table of simulated intensity in machine-readable format in Table \ref{tab:radex_grid}.

\bibliography{references}{}
\bibliographystyle{aasjournal}



\end{document}